%% file: main.tex
\documentclass[compsoc,conference,a4paper,10pt,times]{IEEEtran}
\IEEEoverridecommandlockouts

\usepackage{soul}
\usepackage[table]{xcolor}  %
\usepackage{subcaption}

\usepackage{xurl}
\usepackage{tcolorbox}
\usepackage{amssymb}
\usepackage{xspace}
\usepackage{multirow}
\usepackage{makecell}
\usepackage{amsmath}

\usepackage{hyperref}

\newcommand{\sysname}{XNET\xspace}

\usepackage{tikz}
\usetikzlibrary{shapes.geometric}

\usepackage{listings}

\definecolor{lightgray}{gray}{0.9}

\begin{document}

\title{XNET: Intelligent Dynamic Sampling for High-Speed Network Security Monitoring}
\author{
Thomas Papastergiou$^{1}$,
Karthika Subramani$^{1}$,
Joseph Reilly$^{1}$,\\
Boladji Vinny Adjibi$^{1}$,
Pierros-Christos Skafidas$^{1}$,\\
Roberto Perdisci$^{2}$,
Manos Antonakakis$^{1}$\\[1ex]
$^{1}$Georgia Institute of Technology \quad
$^{2}$University of Georgia\\[0.5ex]
{\small\texttt{\{tpapastergiou, ksubramani, jreilly38, sadjibi3, pskafidas3, manos\}@gatech.edu}}\\
{\small\texttt{perdisci@uga.edu}}
}

\maketitle

\input{abstract}

\input{introduction}

\input{background}

\input{sys_overview}

\input{traffic_tiers_policies}

\input{methodology}

\input{evaluation}
\input{discussion}

\input{related_work}

\input{conclusion}

\bibliographystyle{IEEEtran}
\bibliography{reference}

\input{appendix}

\end{document}

%% file: abstract.tex
\begin{abstract}

Growing network speeds, with 100GbE line rates becoming common in modern enterprise networks, pose challenges to operators and security applications, as they struggle to scale their operational efficiency accordingly, without relying on costly hardware, excessive sampling, or complex distributed deployments. Unintentional loss due to stochastic packet sampling often produces low-quality traffic, further risking missed detection of critical security incidents, particularly those hidden in typically low-rate traffic, such as APT/malware command-and-control communications.

In this paper, we introduce \sysname, a system that monitors traffic at line rate using commodity hardware and applies dynamic sampling to amplify the visibility of high security value traffic. \sysname leverages Linux's XDP technology to process packets efficiently, classify them based on their security value, and sample them as per configured policies. The outcome is a reduced packet stream in which the security-relevant portion of the traffic is amplified at the expense of less interesting traffic segments. \sysname is a highly flexible, scalable and dynamic system that can be adapted based on a network's needs. 
We deployed \sysname in a large real-world network using only commodity hardware, where our results show that \sysname can achieve up to 84\% traffic reduction with no packet loss while increasing the visibility of otherwise negligible traffic fivefold. With controlled stress tests, we further demonstrate \sysname's scalability up to 100Gbps. Additionally, we show that XNET sampling led to a detection rate of 99.6\% in an IDS application.
\end{abstract}

%% file: introduction.tex
\section{Introduction}
\label{sec:intro}

As enterprise networks' connectivity speeds continue to increase, with 100+ GbE Internet links becoming common, network traffic analysis and security applications have struggled to keep up~\cite{HU2020IDS100Gbps}. Even with costly specialized hardware (e.g., FPGA-based network  cards~\cite{napatech,Zhao2020IPS100Gbps}), which minimize packet loss at the interface and kernel level, upstream security applications, such as network intrusion detection systems (NIDS), network audit logging, and others, can suffer from excessive resource consumption (CPU, memory, etc.) and thus incur significant packet loss, which translates into a loss of visibility over high-value, security-related network events. For instance, a NIDS that receives significantly more packets per second than it can process will naturally have to discard many of them \cite{Alikhanov2022,brauckhoff2006impact,mai2006sampled,Hadian2017Sampling}, thus potentially missing critical security events, especially in case of low packet-rate traffic, such as malware/APT command-and-control (C2) communications, targeted remote exploitation attempts, or stealthy data exfiltration~\cite{bilge2012before, nadler2019detection}. 

In this paper, we propose \sysname, a {\em security-aware dynamic traffic categorization and sampling system} that aims to efficiently identify {\em high-priority network traffic} at line rate and to {\em reduce} the number of packets with low-security value that reach upstream enterprise network security applications. Specifically, \sysname focuses on increasing visibility over the highest interest traffic for target upstream security applications such as NIDS (e.g., Suricata~\cite{suricata}) and packet storage and audit logging systems (e.g., Arkime~\cite{arkime}). These security applications parse each packet they receive, thus consuming precious CPU cycles even on packets that have low security value, such as encrypted traffic towards benign web services. \sysname's main goal is to intelligently reduce the volume of traffic with low security value that reaches these upstream security applications, while preserving traffic that is more likely to carry observable malicious activities. In other words, \sysname aims to act as a traffic filter that categorizes traffic in real-time to magnify visibility over higher-risk traffic, doing so at line rate with near zero packet loss and using only commodity hardware.

To understand \sysname's practical benefits, consider a hypothetical NIDS appliance that is deployed at the edge of an enterprise network and that can process up to 20Gbps of traffic. If network traffic volume exceeds the NIDS' capacity, its visibility over potentially malicious traffic will start to decrease significantly. As traffic approaches 40Gbps, each packet will have only a 50\% chance of being processed. Worse yet, packets related to the same traffic flows will be heavily sampled at random, thus preventing proper flow reconstruction and undermining the ability to identify malicious activities. One possible option is to scale the network monitoring infrastructure by investing in additional specialized hardware, such as one or more additional NIDS appliances and higher-scale network packet brokers, which can be quite expensive~\cite{gigamon}. \sysname aims to solve the NIDS visibility issue at a much lower cost, by reducing the volume of traffic that reaches the NIDS to a sustainable level while preserving visibility over high-risk traffic. In practice, \sysname acts as an intelligent traffic prioritization filter that can be placed in front of the NIDS and can run on low-cost commodity hardware to categorize traffic at line rate and only pass high-value traffic to the upstream security application.

\sysname's design is based on the following insights:

\begin{itemize}
\item  {\textbf{Forced Indiscriminate Sampling:}} Security applications that cannot keep up with traffic speeds crossing an enterprise network's perimeter will be forced to lose packets, typically in a randomized fashion determined by stochastic packet arrival times. Effectively, this is equivalent to applying random sampling indiscriminately across all incoming traffic, which can result in significant loss of visibility over ``small'' flows~\cite{ramachandran2008fast} and ``rare'' traffic, namely traffic towards rarely seen hosts and flows that carry a relatively small number of packets.
\item {\textbf{Low-Rate Attacks:}} Critical security events, such as malware/APT C2 communications, exploit attempts, etc., typically consist of low-rate traffic to/from ``rare destinations'', namely domain names or destination IP addresses previously never (or rarely) contacted by the enterprise machines.
\item {\textbf{Encrypted ``Noise'':}} A significant fraction of enterprise network traffic consists of encrypted traffic towards popular web services (e.g., web search, social media, video streaming, etc.). Since the vast majority of this ``bandwidth-clogging'' traffic is highly likely benign and cannot be easily inspected due to the pervasive use of encryption, it generally holds little value for NIDS and network audit logging systems, while potentially consuming substantial NIDS processing time and audit logging storage.
\end{itemize}

As the majority of network traffic is directed toward popular services, attack detection is akin to ``finding a needle in a haystack'', with previous works~\cite{bilge2012before,Ruth2022WWW} indicating that zero-day attacks are generally targeted and affect only a limited number of hosts, typically involving endpoints rarely used or previously unseen within the enterprise network. Therefore, our approach aims to magnify visibility over traffic towards ``rare destinations'', which is often responsible for concealing malware/APT communications or targeted compromise attempts, by favoring such type of traffic while reducing traffic that is more likely benign.

We design \sysname to support high traffic speeds (e.g., 40-100 Gbps) with virtually no packet loss on commodity hardware by leveraging the XDP feature of Linux kernels. XDP ({\em eXpress Data Path}) ~\cite{hoiland2018express} is a component of recent Linux kernels that enables custom packet processing at the network driver's level, bypassing the Linux network stack. To implement \sysname, we developed an XDP program that categorizes network packets early, immediately after the network card receives them, and applies different (dynamic) sampling policies based on the packets' category. The XDP program then decides which packets will be preserved and passed to user space, from where they can be stored or forwarded to an upstream application.

\sysname categorizes traffic using a combination of external data sources and local network behavior. It incorporates global domain popularity rankings, such as Tranco~\cite{pochat2018tranco}, to identify widely accessed destinations and pairing this with observations from within the enterprise network. For example, a domain or IP address that is frequently contacted by many enterprise clients, but not globally popular, is unlikely to be associated with early-stage malware/APT c2 activity. Traffic to such destinations can therefore be sampled more aggressively. At the same time, \sysname continuously monitors traffic to detect new or rarely accessed destinations, those that few or no clients have previously contacted, and ensures that this traffic is retained. This feedback loop allows \sysname’s traffic categorization and sampling policies to evolve based on actual network activity. The system is also designed to be flexible, giving administrators control over how sampling policies are assigned to different traffic categories.

To evaluate \sysname, we perform both stress test experiments in a lab setting and performance measurements in a real-world deployment on a 100 Gbps link connecting a large academic network to the Internet. We also show that \sysname greatly reduces the load on upstream security applications (e.g., Suricata~\cite{suricata} and Snort \cite{snort}) by 78\%-84\% through amplifying visibility on malware-related network traffic.

In summary, we make the following key contributions:
\begin{itemize}
    \item We design and develop \sysname, a security-aware dynamic network traffic categorization and sampling system to process high-speed traffic at line rates up to 100Gbps using only commodity hardware.
    \item We deploy \sysname on live traffic mirroring a 100GbE link with traffic speeds reaching up to 87Gbps. Through applying dynamic sampling by leveraging knowledge inferred from local network behaviors and external intelligence, \sysname was able to achieve a drastic traffic reduction of 78\%-84\% while retaining traffic of high security value.
    \item We conduct an ablation study of \sysname's XDP components to measure the performance overhead of its packet processing operations at various loads and demonstrate its capability to scale up to 100Gbps.
    \item We measure \sysname's influence on IDS applications, by showing the direct correlation between increased visibility into security-focused traffic and the alerts generated, resulting in {\em99.6\%} detection rate.
    
\end{itemize}

%% file: background.tex
\section{Background}

\subsection{Challenges in Network Traffic Monitoring}

The high-speed nature of modern backbone links, with traffic reaching hundreds of Gb/s, poses practical constraints on the CPU and memory resources available in network equipment, forcing network operators to balance between performance and accuracy. Current approaches are caught between two extremes: low-fidelity general-purpose methods like random packet sampling, which are computationally feasible but fail to capture detailed network behaviors, and high-fidelity, complex algorithms that offer precision at the cost of excessive resource consumption, translating into heavy packet loss. This trade-off between precision and practicality remains a key challenge in designing and implementing effective network monitoring solutions.

Traditional packet sampling, the most common of which is random packet sampling, inherently biases the data toward larger flows or ``heavy hitters,'' as the likelihood of sampling a flow is proportional to its size and the probability of recording network communications with a given endpoint is related to its popularity. While this makes it effective for recording and meaningfully analyzing samples of high-volume flows (e.g., for media streaming) and popular network endpoints, it significantly degrades visibility over rare events, such as targeted attacks or malware/APT communications, which often manifest through small flows towards rarely (or never before) observed endpoints.  Likewise, alternative methods such as ``smart sampling''~\cite{duffield2002properties} and ``sample-and-hold''~\cite{estan2003new} are better in estimating heavy hitters but tend to distort the original traffic distribution, resulting in a high rate of false positives in security applications~\cite{Alikhanov2022,brauckhoff2006impact,mai2006sampled,Hadian2017Sampling}.

Alternatively, flow aggregation techniques like NetFlow~\cite{netflow} and sFlow~\cite{sFlow} are widely used for coarse-grained traffic analysis because of their reduced storage footprint. These methods aggregate sampled packets into flow records, which are useful for general metrics like total traffic volume. However, they have difficulty producing high-fidelity data \cite{duffield2003estimating,ramachandran2008fast} unless operated at high sampling rates, which increases resource consumption. In fact, it is not uncommon for large networks to use packet sampling rates of 1:1000 or higher~\cite{fastnetmon,cloudflare_samplingrate, brauckhoff2006impact}. Additionally, as mentioned earlier, they often miss details about small and medium flows, making them less suitable for security applications~\cite{li2016flowradar,estan2004building,carela2011analysis}, where finer granularity is needed.

\input{xdp_dpdk}

%% file: xdp_dpdk.tex
\subsection{Accelerated Packet Processing Frameworks}
\label{xdp_overview}

Accelerated packet processing frameworks like XDP (eXpress Data Path) and DPDK (Data Plane Development Kit) offer optimized software solutions for performance and scalability to meet the demands of real-time packet filtering, analysis, and manipulation. XDP \cite{hoiland2018express} is a high-performance, low-level networking feature introduced with Linux kernel 4.8. It serves as an early hook point in the kernel, which enables attached eBPF programs to perform ultra-fast custom packet processing directly at the network interface card (NIC) driver level, bypassing most of the operating system networking stack. For details around eBPF we refer the reader to \cite{ebpf}.

As an XDP program processes each packet independently, it needs a mechanism to maintain information across multiple packets. eBPF maps allow retaining state information, such as counters or connection data, which would otherwise be lost between packet events. eBPF maps are data structures that come in various types, like arrays, hash maps, queues, per-CPU arrays, and LRU (Least Recently Used) hash maps, each suited to different needs. This flexibility allows XDP programs to manage packet counts, IP addresses, protocol statistics, and other custom metrics tailored to specific network policies. Finally, eBPF maps are accessible to both kernel space (where XDP runs) and user space. This dual access enables XDP programs to share data with monitoring and management tools, allowing for live updates, retrieval of statistics, and even passing packets efficiently to the user space through the specialized {AF\_XDP} sockets. We refer the reader to Appendix \ref{app:design_choices} for a comparison with DPDK and the reasons we opted to use XDP.

%% file: sys_overview.tex
\begin{figure}[t]
  \includegraphics[width=0.95\columnwidth]{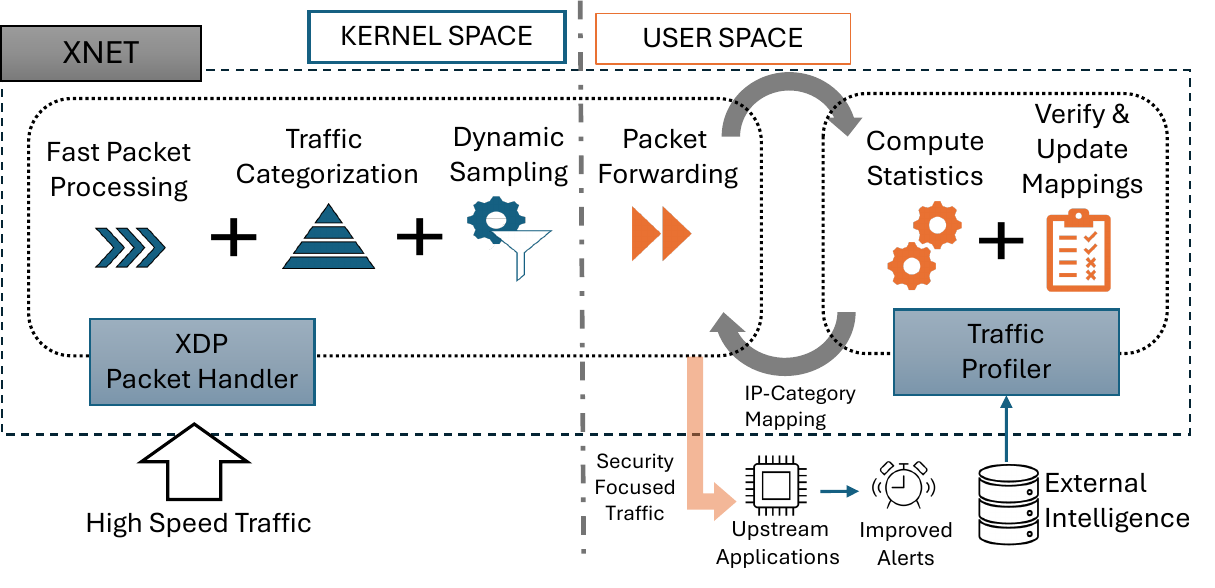}
  \caption{High-level overview of XNET.}
  \label{fig:sysoveview}
\end{figure}

\section{\sysname Overview}

We designed \sysname to {\em aid high-speed network security applications} such as NIDS, packet storage and indexing, etc., achieve significantly improved visibility over high-value security traffic using only commodity hardware. Importantly, \sysname \textit{is not itself a detection system}. Rather, it is a {\em traffic prioritization} system in support of third-party network security applications. 

Figure~\ref{fig:sysoveview} depicts a simplified version of our system that ingests a stream of high-speed traffic, leverages global and local network intelligence to identify network behavior profiles and usage patterns, and applies this intelligence to efficiently partition traffic flows into multiple categories. Each traffic category is mapped to a sampling policy, which determines \textit{whether} and \textit{how} traffic belonging to a given category should be reduced. The intended outcome is a significantly reduced packet stream that undersamples traffic related to heavy-hitter flows and highly popular destinations, while magnifying visibility on ``low-and-slow'' traffic for upstream security applications.

\subsection{Design Goals and System Components}
\label{sec:design_goals}
\sysname's architecture includes two main subsystems: (i) the XDP Packet Handler and (ii) the Traffic Profiler, which can be easily customized and extended to adapt to a specific network's requirements. Their design goals are as follows:

\newcommand{\rectangledDG}[1]{
  \tiny
  \tikz[baseline=(char.base)]\node[draw,rectangle,rounded corners,inner sep=2pt](char){DG#1};
}

\begin{enumerate}
    \item Achieve significant traffic reduction to prevent upstream applications from being overwhelmed and forced to indiscriminately (i.e., randomly) sample incoming network packets.
    \item Adopt a dynamic sampling approach that accounts for the traffic category.
    \item Develop a configurable and extensible system that offers flexibility to network administrators.
    \item Ensure that the system is scalable and performs at a close-to-line rate using only readily available commodity hardware, without the need for specialized equipment.
    \item Maintain backward compatibility and easy integration with existing upstream applications by being able to efficiently forward the dynamically sampled traffic to a configurable destination network interface.
\end{enumerate}

\vspace{2pt}
\noindent \textbf{XDP Packet Handler}: We utilize Linux's eXpress Data Path (XDP) framework to build a custom distributed (multi-core) XDP program, achieving line-rate packet processing directly in the kernel. Our multi-core processing module efficiently load-balances raw packets across CPU cores, categorizes traffic, applies category-specific sampling policies, and then forwards sampled packets to an upstream application for further analysis. Our XDP program comprises a kernel-space and a user-space component, which communicate via eBPF maps~\cite{ebpf}. In the kernel, we extract header information for each packet and consult the eBPF maps to categorize it, apply the appropriate sampling policy, and either forward it to the user space or drop it. Packet forwarding is achieved via {AF\_XDP} sockets~\cite{afxdp}, which bypass most of the networking stack, allowing packet transfer with minimal latency and CPU usage to a shared memory location. In user space, the XDP Packet Handler offers multiple data ingestion options that accommodate the most plausible operational scenarios, such as PCAP storage, forwarding 
via a separate interface, or direct integration through hook points in the code.

\vspace{2pt}
\noindent \textbf{Traffic Profiler}: While the XDP Packet Handler is responsible for filtering and collecting traffic, it relies on dynamically computed intelligence to determine the traffic categories and apply the sampling policies accordingly. The Traffic Profiler component processes
the volume of traffic forwarded by the XDP Packet Handler and its primary responsibility is to compute traffic statistics and build a behavior profile for the external network endpoints. For instance, we compute the number of clients internal to the deployment network that in the recent past have communicated with a given external destination IP address, which can indicate whether this IP is locally ``popular.'' Likewise, we track external ``heavy-hitter'' IPs that tend to exchange very large amounts of traffic with network clients. Ultimately, this module produces intelligence in {\em IP-to-Category} mapping, where IP refers to an external IP that's not part of the local network and {\em Category} refers to a predefined set of traffic categories. After its initial deployment, the Traffic Profiler continuously accumulates and updates the required intelligence at frequent intervals by observing the traffic crossing the network perimeter. This intelligence is then communicated to the kernel component by updating the eBPF map with the IP-Category mappings, forming a feedback loop that enables instant policy updates.

%% file: traffic_tiers_policies.tex
\subsection{Traffic Categories and Sampling Policies}
\label{sec:traffic_tiers}

Next, we describe the sampling policies and traffic categories that serve as a reference throughout the paper. \sysname is deliberately designed to be easily configurable, supporting the addition and tuning of traffic categories and policies with minimal development effort (see discussion in Section~\ref{subsec:extensions}). In defining the sampling strategies and traffic classifications, we follow these guiding principles:

\begin{itemize}
  \item {\em Traffic visibility}: 
  One of our main design goals is to reduce the volume of traffic forwarded to upstream applications to prevent them from becoming overwhelmed and resorting to indiscriminate (i.e., randomly) sampling of incoming packets. Simultaneously, we aim to fully preserve traffic towards ``rare'  destinations, which could reveal critical security events, such as malware/APT C2 communications, exploit attempts, etc. Thus, traffic categories and related sampling policies should be identified based on a trade-off between traffic volume reduction and the likelihood that the upstream applications will benefit from the filtered traffic.
  
  \item {\em Efficiency}: Traffic categorization should rely on simple packet header analysis to sustain high-speed packet arrivals. For instance, we should avoid parsing complex protocol payloads (e.g., application layer messages) and instead focus on basic features, such as IP addresses and TCP/UDP ports, to identify external endpoints and distinguish between categories.
\end{itemize}

Below, we first introduce the different traffic sampling policies that \sysname implements and then define the traffic categories and how the sampling policies map to them.

\subsubsection{Sampling Policies}
\label{sec:sampling_policies}
The XDP Packet Handler supports a set of sampling policies that are applied at line rate for each packet. 

\vspace{2pt}
\noindent \textbf{Random Sampling}: A packet has a chance to be dropped according to a specified sampling rate. Different sampling rates can be assigned based on the traffic category, ensuring less relevant traffic is sampled more heavily to reduce load.

\vspace{2pt}
\noindent \textbf{First $N$ Packets per Flow}: The XDP program can track active flows using protocol, IP, and port information extracted from packet headers as the identifier. A counter is associated with each flow to record the number of packets seen so far. Once this counter reaches a defined threshold, $N$, subsequent packets in the flow are dropped. This approach retains initial highly informative flow data (e.g., first HTTP request/response headers, TLS Client Hello and the servers' SSL certificate, etc.) while reducing the load from large flows.

\vspace{2pt}
\noindent \textbf{No sampling}: All packets for specific flows are forwarded without loss. Reserved for high-priority or sensitive traffic categories where full visibility is essential.

\subsubsection{Traffic Categories}
\label{sec:traffic_categories}
\label{subsec:sampling_category_mapping}

We now describe the traffic categories used as a reference in our evaluation. In addition, inspired by threat assessment systems~\cite{PANRiskLevel}, we assign a security importance level to each traffic category and map them to suitable sampling policies accordingly. Consider these as mere examples of categories and mappings that a network administrator can use to demonstrate how \sysname can be used in practice, but we emphasize that additional categories can be integrated into the system with limited effort.

 \vspace{3pt}
\noindent \textbf{DNS}: DNS packets are treated as a distinct category that we capture in its entirety. Though representing a small portion of overall traffic, DNS data is provenly critical for monitoring and security~ \cite{antonakakis2010building,antonakakis2011detecting,nadler2019detection,perdisci2020iotfinder} and a No Sampling policy is applied in this case.

\vspace{3pt}
\noindent \textbf{Rare:} This category is reserved for ``low and slow" traffic that upstream security applications may more likely overlook if they are forced to randomly sample packets due to high packet arrival speeds. Such traffic often harbors malicious activities (e.g., APTs, ransomware, low-rate DoS attacks, etc.~\cite{wang2024lordma,Shan2017TailAttacks}). We label as ``Rare'' all traffic associated with external IPs that only a few (or no) clients have recently contacted. This traffic portion has a higher likelihood of containing attack traffic, hence we classify it as "high risk" and apply the No Sampling policy, ensuring that all packets are forwarded to the security application for comprehensive analysis. 

\vspace{3pt}
\noindent \textbf{Shared Services}: This category includes traffic towards shared network infrastructure, such as cloud platforms, virtual hosting, and content delivery networks (CDNs). Since many popular services rely on these shared infrastructures for their operations, we identify traffic associated with such services by regularly collecting
public IP range data from providers’ websites. Although traffic towards popular CDNs and cloud IP providers is for the most part benign, adversaries can abuse the related ``for rent'' IP space to host malicious network infrastructure. We therefore consider this traffic as ``medium risk'' and apply the ``First N Packets per Flow'' sampling policy with a tunable threshold $N_s$ (see Section~\ref{sec:threshold_selection}). For instance, this allows us to capture the first (unencrypted) packets for new TLS sessions, which can provide useful information such as the server name indication (SNI), the cipher suites available to the client, etc.

\vspace{3pt}
\noindent \textbf{Locally Popular}: This category represents traffic towards external destinations with significant relevance within the local network. Local popularity is determined by the number of unique internal clients\footnote{The number of clients may represent a conservative estimate of the true extent of interactions, as some internal IP addresses might correspond to multiple users or systems due to specific network configurations (e.g., due to network address translation in internal private subnets)} that have communicated with a particular external IP within a moving time window and can change over time. IPs linked to shared services are excluded from this category to avoid confusion, as those endpoints may serve both popular and less popular services. This traffic tends to be overwhelmingly of benign nature. We therefore assign a ``low risk'' level and apply the Random Sampling policy with sampling rate $\tau_l$, so that only a predetermined fraction of packets is retained, thus greatly reducing traffic volume. 

\begin{figure}[t]
    \centering
    \captionsetup{skip=2pt}
  \includegraphics[width=0.75\columnwidth]{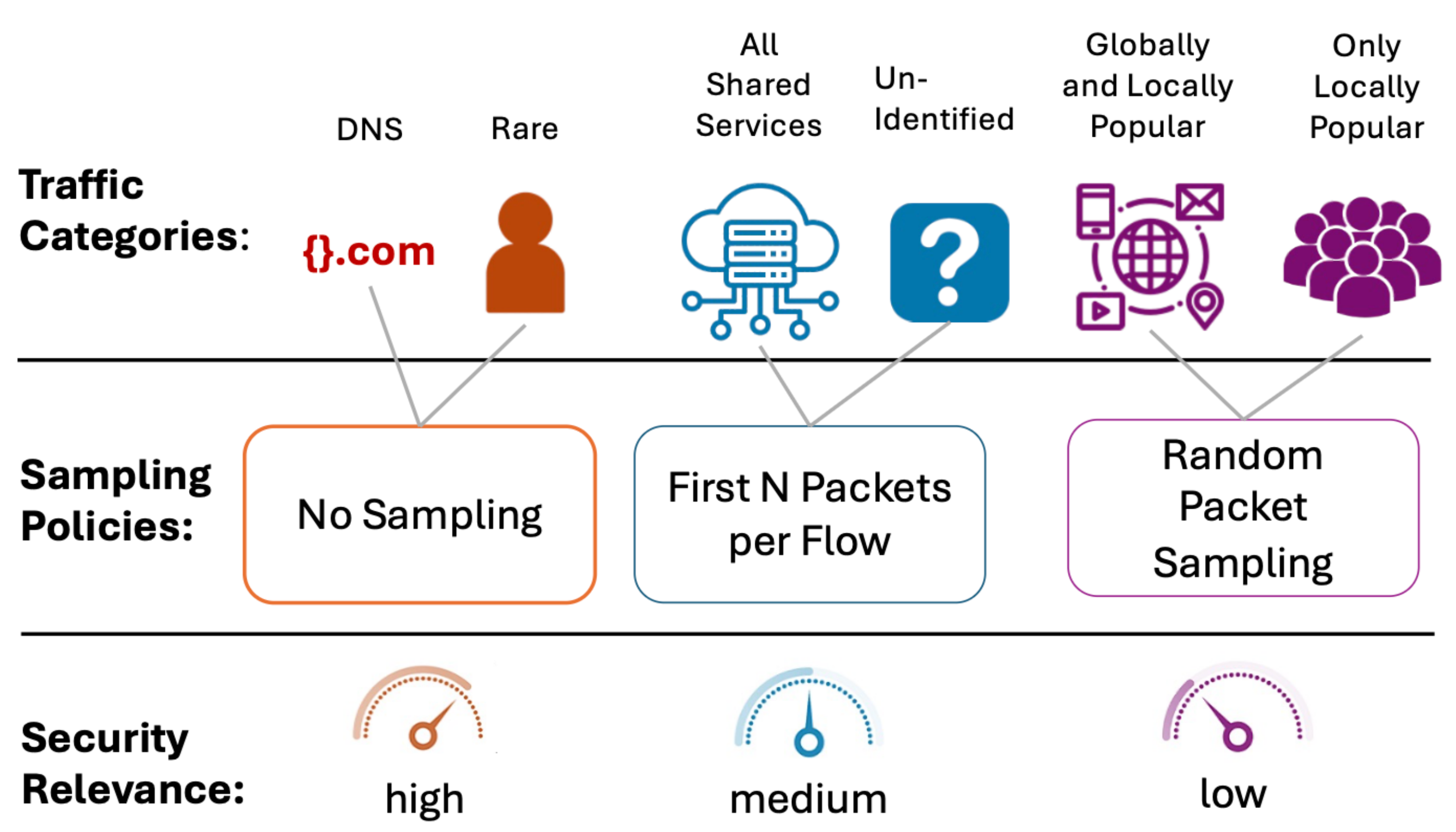}
  \caption{Mapping of Sampling Policies to Traffic Categories. }
  
  \label{fig:mapping_sampling_categories}
\end{figure}

\vspace{3pt}
\noindent \textbf{Locally and Globally Popular:} This category encompasses traffic towards external IPs that are not only flagged as locally popular but also associated with major web services, like Google, YouTube, and other social media or streaming websites~\cite{Ruth2022WWW}. To label such external IPs, we leverage passive DNS data to obtain the domain associated with each external IP address observed in the monitored traffic (notice that visibility above the local DNS resolver is sufficient for this purpose). We then evaluate its global ranking by checking the effective second-level domain (eSLD) against widely adopted Tranco~\cite{pochat2018tranco} domain popularity list and apply a tunable ranking threshold (discussed in Section~\ref{sec:threshold_selection}). This category is also assigned a ``low risk'' threat level, and we apply a more selective Random Sampling rate $\tau_g < \tau_l$, to further reduce unnecessary traffic. More notably, this category excludes traffic belonging to shared services.

XNET continuously monitors DNS traffic and maintains up-to-date mappings between Tranco domains and their observed resolved IPs, because while the top domains in the Tranco list do not change very frequently, their resolved IPs may change often.

\vspace{3pt}
\noindent \textbf{Unidentified}: We label ``Unidentified" traffic that doesn’t fit into any of the previous categories. It includes  traffic to external endpoints that do not qualify as either Locally Popular or Rare, but are still contacted by a moderate number of clients. We assign this traffic a ”medium risk” level and apply the ``First N Packets per Flow'' sampling policy with a threshold $N_u$, similar to Shared Services.

The presented mapping between sampling policies and traffic categories is flexible, and \sysname allows administrators to adjust these assignments to suit specific security requirements and priorities.

%% file: methodology.tex
\section{\sysname System Details}
\label{sec:sys_details}

This section offers a closer look at \sysname and its two primary, interconnected subsystems: the XDP Packet Handler (Section \ref{subsec:xdp}) and the Traffic Profiler (Section \ref{subsec:assessor}). The XDP Packet Handler captures raw network packets, assigns them to traffic categories, applies filtering operations, and forwards them to upstream applications. The Traffic Profiler then generates high-level insights from the filtered packets, continuously updating IP-to-category mappings and creating a feedback loop with the Packet Handler.

\input{xdp_packet_handler}

\subsection{Traffic Profiler}
\label{subsec:assessor}

Packets that pass the sampling process are forwarded to the Traffic Profiler (Figure~\ref{fig:traffic_profiler}).
The Traffic Profiler then analyzes the incoming traffic, computes the up-to-date network intelligence, and circulates it back to the XDP Packet Handler, as explained below. 

\vspace{3pt}

\noindent \textbf{DNS Parsing and Flow Extraction}: 
This multi-threaded module condenses incoming traffic and retains only information required for computing traffic statistics. For DNS packets, it specifically parses DNS responses to map queried domains to their corresponding resolved IP addresses, which is needed to identify IPs related to globally popular web services. For non-DNS packets, it analyzes packet headers to extract 5-tuples, {\tt <src\_IP, dst\_IP, src\_port, dst\_port, protocol>}, that identify uni-directional flows. To efficiently aggregate information regarding both directions of a network flow (e.g., total number of packets exchanged between client and server), we use a simple approach whereby we swap source and destination information extracted from packet headers such that the destination IP is always an external IP (we identify external IPs as those that do not belong to the set of pre-defined IPs allocated to the deployment network). Each thread stores information about the flows it encounters on a per-thread hash map.

Next, statistics are consolidated across the packet processing threads. For each flow, we consider its external IP address and compute the number of distinct local hosts that communicated with that IP and the total number of packets exchanged, storing that information in a hash map, called {\tt EXTERNAL\_IP\_STATS}, that stores {\tt <external\_IP, (hosts\_count,packets\_count)>} pairs. To solve the challenging task of estimating in a highly efficient way the number of unique local hosts that communicated with an external IP, we leverage a Bloom filter to track the {\tt <external\_IP, host\_IP>} pairs seen for the current day. For each such pair, we check its presence in the Bloom filter; if it’s not found, we update the {\tt external\_IP}'s count in {\tt EXTERNAL\_IP\_STATS}, otherwise we have already counted this pair and we can ignore it. Given the large number of IP pairs we may see over the course of a day, we opted for a Bloom filter because it enables rapid lookups using constant memory with minimal false positives
In the rare case of a false positive, the only effect is a slightly lower count of local hosts observed for a given external IP. This means that XNET may cause a few external IP addresses to be incorrectly categorized as ``Rare.'' However, this simply leads to a very small increase (as a percentage of all traffic) in the number of packets that will be forwarded to the upstream application, and does not affect the collection of more security-relevant traffic that would otherwise be collected with a deterministic lookup, instead of using Bloom filters. To compute traffic statistics, we maintain a daily record of {\tt EXTERNAL\_IP\_STATS} map for further processing in the next step. The Bloom filter and map are then reset at the end of the day to compute statistics pertaining only to the current day.

\vspace{3pt}

\noindent \textbf{Extracting IP-Based Intelligence:} To ensure Traffic Profiler has sufficient information for mapping external IPs to traffic categories, we first consolidate all required parameters. Accordingly, we incorporate the previous day's IP statistics along with IP statistics collected from the current day's rolling time window, so that at least a 24-hour time frame is considered at any given moment. Next, these statistics are combined with data from external feeds, such as domain-based popularity lists (Tranco) and IP ranges scraped from published lists of popular CDN and Cloud services. The gathered parameters include the count of local hosts and the number of packets derived from flow statistics, in addition to any ranking or connection to shared services obtained from external feeds.

As a final step, the Traffic Profiler periodically verifies if the existing IP-Category mapping still holds after the recently updated parameters or needs reevaluation. Let's consider an IP that, in the current day of traffic, no longer satisfies its previously assigned category's criteria. In this case, we change the IP's category to the new one only after an ``IP retention period,'' which represents a predetermined time frame the IP must remain in its current category before it can be reclassified. The retention period ensures stability by avoiding premature reassignments and unnecessary updates due to temporary changes in network traffic behaviors. 

We compute the category retention period by calculating the difference in days between the current date and the last seen date recorded for a given IP-Category pair.
Once the retention period expires, the IP can be assigned to the category computed on the current last day of traffic. Conversely, if an IP continues to satisfy all criteria for its previously assigned category, its ``last seen'' record for the category is updated to the current date. Any update to the IP-Category mapping maintained by the Traffic Profiler immediately triggers an eBPF map update so that the XDP Packet Handler can make use of the new mappings during real-time packet processing.

%% file: xdp_packet_handler.tex
\subsection{XDP Packet Handler}
\label{subsec:xdp}

The various subsystems of the XDP Packet Handler are shown in Figure \ref{fig:packet_screening}. When a packet arrives, the NIC places it into a receive (RX) queue. Modern NICs support multiple RX queues and, using technologies like Receive Side Scaling (RSS), can distribute network traffic across multiple CPU cores to optimize processing. Tools like Linux’s \textit{ethtool} allow for tuning the number of RX queues, up to the maximum number supported by the hardware. Once packets populate an RX queue, the NIC generates an IRQ interrupt specific to that queue, triggering the NIC's device driver. By loading our XDP program at the driver-level hook, we intercept packets at an early stage and process them in a highly parallelized way, using a number of instances equal to the number of RX queues.

The XDP Packet Handler performs three key tasks: it categorizes the network traffic, applies the related sampling policy (see Section~\ref{sec:traffic_tiers}) and forwards the filtered traffic to upstream applications.

\subsubsection{Implementation of Traffic Categorization}
\label{subsec:packet_screening}
\begin{figure}[t!]
\centering
\captionsetup{skip=2pt}
  \includegraphics[width=0.8\columnwidth]{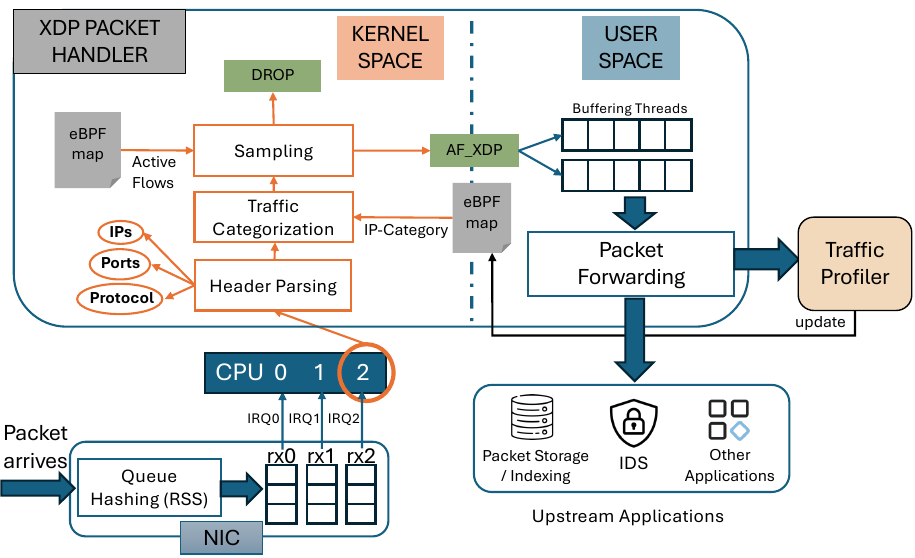}
  \caption{Packet processing in the XDP Packet Handler.}
  \label{fig:packet_screening}
\end{figure}

The first step is determining the traffic category a packet belongs to, so that the corresponding sampling policy can be applied. Figure \ref{fig:packet_screening} illustrates the packet's flow through the various stages in the XDP program. First, every packet is parsed in the kernel space, and necessary high-level information from the packet headers, such as IP addresses, protocols, and, when applicable, ports, is extracted. DNS queries and responses are efficiently identified by looking for UDP packets with port 53 in the destination or source port fields, respectively. Note that port 53 remains widely used for DNS in operational networks. Even when encrypted DNS (DoH/DoT) is employed, the TLS ClientHello typically exposes the SNI, which is captured by XNET’s first-N-packet policy and still enables effective domain–IP association.
For non-DNS traffic, we consult an eBPF hash map (curated by the Traffic Profiler) where the key is an external IP address (the source or destination IP address of the observed packet) and the value is the traffic category associated to that address (an introduction to eBPF maps can be found in Section \ref{xdp_overview}). Packets that cannot be matched to a category are classified as ``Rare'' by default, since the flow relates to a never-before-seen external IP (relative to the time window considered for creating the map). The maximum number of entries in the maps is limited only by the system's memory, effectively enabling categorization across the entire IPv4 space.

\subsubsection{Implementation of Sampling Policies}

 Once the packet is assigned a traffic category, it is sampled accordingly (Section~\ref{subsec:sampling_category_mapping}). Depending on the sampling policy, the packet is either dropped or forwarded to user-space via an \url{AF_XDP} socket. For instance, DNS and Rare traffic are subject to the ``no sampling'' policy and all packets in these categories are thus passed directly to the user-space.

To implement random sampling at the kernel level, we use the helper function \textit{\url{bpf_get_prandom_u32()}}, which generates a random integer that we compare against a threshold to achieve a given target sampling rate. 

For categories that require flow tracking and retention of only the first $N$ packets per flow, such as the Shared Services or Unidentified categories, we use a secondary eBPF hash map of type \url{BPF_MAP_TYPE_PERCPU_LRU_HASH} with the tuple \textit{(srcIP, dstIP, srcPort, dstPort, proto)} as the unique key for each flow. The corresponding value is the packet count for the flow, which increments with each packet or initializes to one if the flow does not already exist.

This map’s configuration as a PERCPU and LRU (Least Recently Used) hash map is deliberate. The PERCPU attribute ensures that each CPU core maintains its own instance of the map, reducing race conditions and aligning well with how RSS (Receive Side Scaling) distributes packets across RX queues based on header information. The LRU functionality automatically removes the least recently used flows as the map reaches capacity, an efficient compromise for flow expiration without the complexity of tracking inactivity or inspecting TCP flags. Effectively, this resets packet counters for the evicted flows. 
The optimal size of the LRU cache depends on network conditions, which tend to vary over time. However, the size can be estimated by using historic traffic statistics and choosing a conservatively large approximation. It is important to notice that even if the chosen LRU cache size were to be smaller than optimal and a flow gets prematurely evicted, visibility is not degraded (missed packets). Instead, once a flow is evicted, the ``first $N$" counter effectively resets to zero and when the flow's next packet arrives, it will be captured and counted as the flow's first packet. Effectively, premature evictions increase visibility by retaining more packets per flow at the expense of the overall traffic reduction rate.

\begin{figure}[!t]
\captionsetup{skip=2pt}
  \includegraphics[width=\columnwidth]{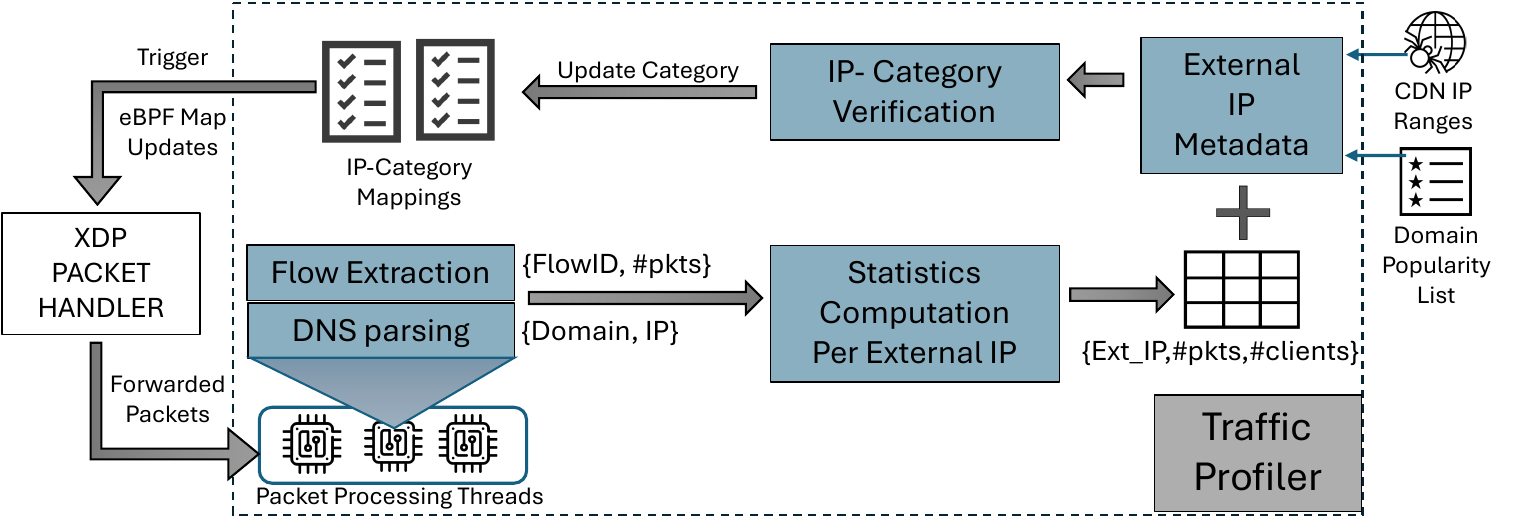}
  \caption{The Traffic Profiler subsystem.}
  \label{fig:traffic_profiler}
\end{figure}

\subsubsection{Forwarding Packets To Upstream Applications}

From there, packets can be accessed by consumer applications through three main options:

\begin{enumerate}
    \item \sysname can save timestamped packets in PCAP format at a user-specified location, enabling upstream applications to process these files independently. Although this method introduces some I/O overhead due to disk writing, its impact remains manageable in our deployment setup even for high packet rates, allowing for modular operation of different system components.
    \item Applications like IDSs or packet brokers can receive packets directly on a dedicated virtual or physical interface. In this setup, \url{AF_XDP} sockets are used to efficiently route packets to the designated interface, providing seamless and configurable integration within operational environments.
    \item For performance-critical applications, a direct callback function can be implemented to process packets immediately as they arrive. This method provides the fastest integration with minimal latency, but requires more programming effort.
\end{enumerate}

To optimize performance for options 1 and 2, we employ multiple user-space threads to handle packets from different RX queues (RXQs). These threads accumulate packets into predefined buffers, which are processed either by writing to a file (in method 1) or sending to an interface (in method 2) once the buffer reaches capacity. This multithreaded buffering helps to manage packet flow and maintain high throughput.

%% file: evaluation.tex
\section{Evaluation Results}
\label{sec:eval}

In this section, we comprehensively evaluate \sysname's performance through both controlled lab experiments and deployment in a large real-world academic network. First, we discuss how we configure the system's parameters (Section \ref{sec:setup}) and present experimental results obtained in the real-world deployment, which show that \sysname can reduce traffic by up to 84\% with no packet loss, when deployed on a live network with external traffic (also referred to as ``north-south'' traffic) reaching up to 87 Gbps (Section \ref{sec:real_deployment}). Next, we conduct an ablation study to understand the performance of \sysname's system components under different traffic conditions and to demonstrate its scalability to 100 GbE links (Section \ref{sec:performance_scaling}). Finally, we assess the security value of the reduced traffic with a case study, using Suricata as the upstream security application (Section \ref{sec:suricata}).

\subsection{Experimental Setup}
\label{sec:setup}
\subsubsection{Hardware}
To perform the experiments, we use the following setup. We have deployed two identical servers, each equipped with two Intel Xeon Gold 5318N CPUs @ 2.10GHz (48 physical cores total across 2 sockets), 250GB of RAM, 20TB of NVMe storage and a dual-port 100GbE NVIDIA Mellanox ConnectX-6 NIC. One of the 100GbE ports is connected directly to the sibling server with a 100GbE link, while the other port receives the live network traffic feed from our institutional network via a 100Gbps fiber connection. Both boxes run Ubuntu Server 22.04 LTS.

\subsubsection{Parameters Selection}
\label{sec:threshold_selection}

In this section, we explain our rationale for choosing threshold values for the parameters used to categorize traffic. These values are selected through empirical analysis of unsampled, real-world traffic collected during our pilot experiments that included data from two different organization's network. However, further experiments were carried out in only one of the two networks. For each external destination $x$, we compute the following metrics:

\begin{enumerate}
    \item the number of local clients that contacted $x$, denoted $LC(x)$ 
    \item the volume of traffic exchanged with $x$, $V(x)$
    \item the global popularity rank $GR(x)$ obtained from passive DNS data and the Tranco list.
    \item an indicator $SS(x)\in\{0,1\}$ specifying whether $x$ belongs to a known Shared–Service IP range
\end{enumerate}

Let $\theta_R$ be the threshold below which an IP is considered rare, $\theta_{LP}$ the threshold above which an IP is locally popular, and $\theta_{GP}$ the rank cutoff for global popularity. The category assigned to $x$, written $Cat(x)$, is thus

{\tiny
\[
Cat(x)=
\begin{cases}
Rare, &  LC(x)\le \theta_R,\\[4pt]
SharedServices, & SS(x)=1  \wedge\ LC(x)\ge \theta_{R},\\[4pt] 
LocGlobPopular, & SS(x)=0 \ \wedge\ LC(x)\ge \theta_{LP}\ \wedge\ GR(x)\le \theta_{GP},\\[4pt]
LocallyPopular, & SS(x)=0 \ \wedge\ LC(x)\ge \theta_{LP}\ \wedge\ GR(x)>\theta_{GP},\\[4pt]
Unidentified, & \text{otherwise}.
\end{cases}
\]}

As we detail the categorization parameters and how the associated metrics are applied, we follow a conservative principle: we prefer retaining more packets and preserving visibility over risking excessive filtering.

\vspace{3pt}
\noindent \textbf{Global Rank Threshold:} 
To determine the upper bound on the popularity rank used to classify a destination~$x$ as \emph{Globally Popular}, we evaluated how the cumulative traffic volume $\sum_{x\,:\,R(x)\le k} V(x)$ grows as each external domain's Tranco rank increases as shown in (Figure~\ref{fig:cdfrankcount}, Appendix). The cumulative curve increases sharply for the highest-ranked domains and then flattens, with roughly $25\%$ of all traffic associated with destinations whose domains satisfy $GR(x) \le 20{,}000$. Extending the cutoff to the top $1$M domains adds only a small additional fraction (around $5\%$), and considering only ranks below $10{,}000$ yields a similar share of about $23\%$. This flattening indicates that the knee of the curve lies near $k \approx 20{,}000$. Accordingly, we adopt $\theta_{GP} = 20{,}000$ as the threshold in our experiments.

\begin{figure}[t]
  \centering
 
  \begin{subfigure}[b]{0.48\columnwidth}
      \centering
      \includegraphics[width=0.95\columnwidth]{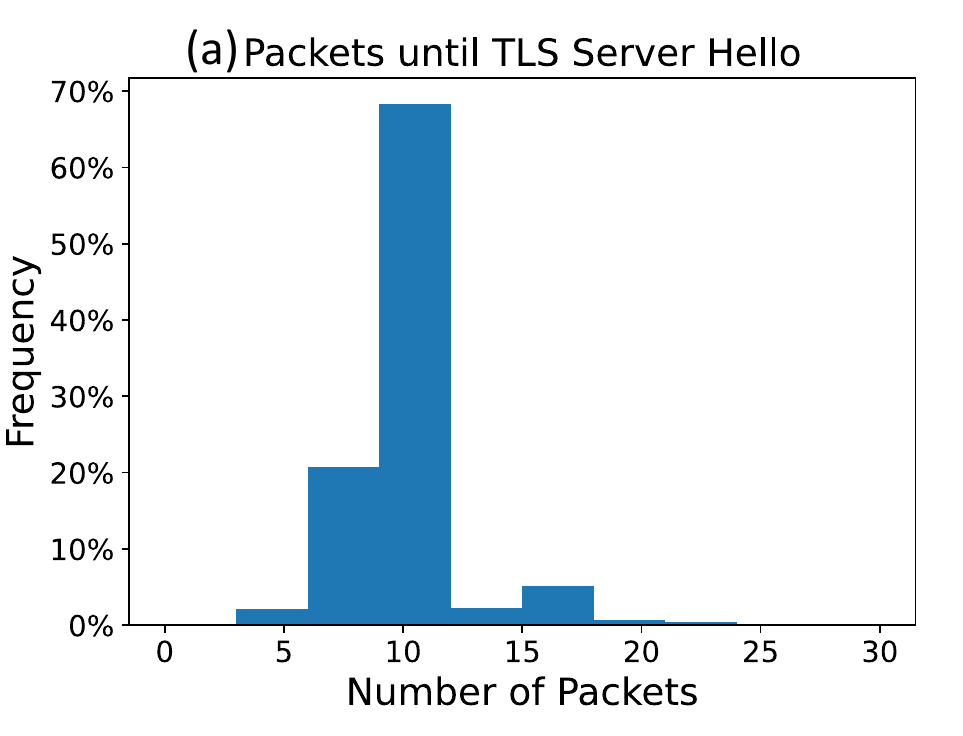}
  \end{subfigure}
  \begin{subfigure}[b]{0.48\columnwidth}
      \centering
      \includegraphics[width=0.95\columnwidth]{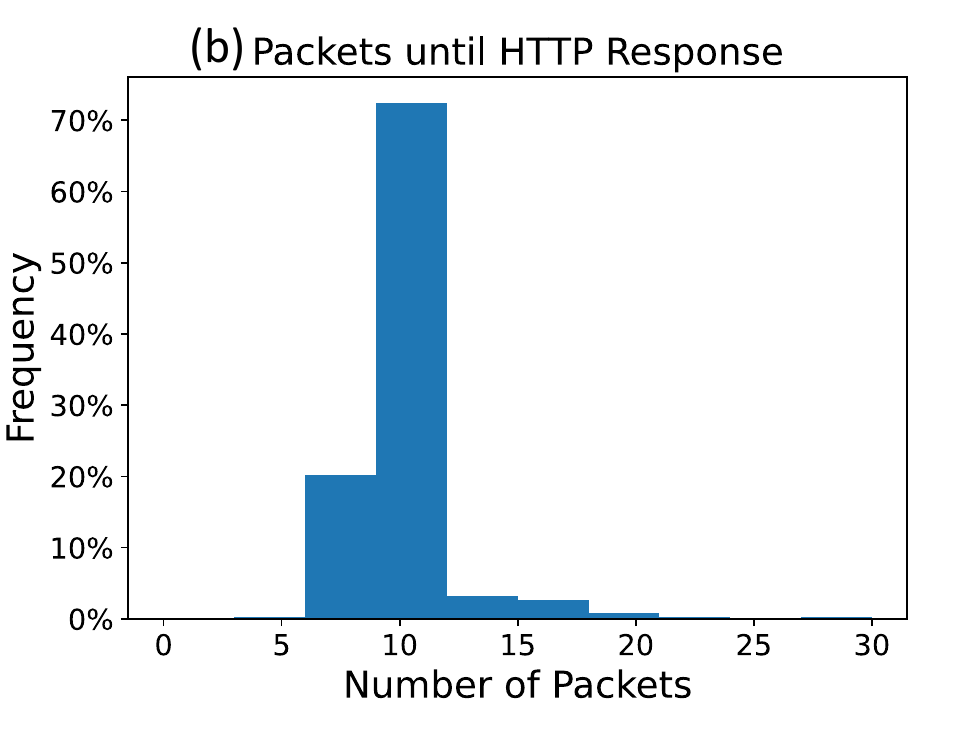}
  \end{subfigure}
  \caption{Distribution of first {\em N} packets associated with a) TLS handshake and b) TCP handshake.}
  \label{fig:fist_n_count}
\end{figure}

\vspace{3pt}
\noindent \textbf{Local Popularity Threshold:} 
To determine how many internal hosts must contact an external destination~$x$ for it to be classified as \emph{Locally Popular}, we examined the distribution of the unique-client metric $LC(x)$ in our pilot dataset. In particular, we focused on destinations associated with highly ranked domains, considering those with $GR(x)\le 1{,}000$, $GR(x)\le 10{,}000$, and $GR(x)\le 20{,}000$. Across these groups, the average number of distinct internal clients remained consistently below $25$ per destination. Motivated by this observation, we select a deliberately conservative threshold of $\theta_{LP} = 100$, meaning that a destination is labeled locally popular only if $LC(x) > 100$. This choice leads \sysname\ to forward more traffic from popular services to the upstream application, reflecting our preference for maintaining visibility in such cases. While this threshold is tunable and can be lowered to achieve greater reduction, we retain the conservative value of $100$ in our experiments to illustrate that \sysname\ achieves substantial traffic reduction even under cautious parameter settings.

\noindent \textbf{Rare Threshold:} 
To select the threshold for the \emph{Rare} category, we rely on the observation that in the early stages of targeted intrusions, such as APT activity or malware infections, only a small number of internal hosts typically initiate contact with external command-and-control or attacker-controlled infrastructure. Because preserving full visibility into such traffic is critical, we adopt a conservative cutoff $\theta_{R} = 15$, and classify a destination~$x$ as rare when $LC(x) \le 15$. Note that even popular destinations were connected to only $25$ clients, on average. Although this threshold may appear relatively high, it reflects our preference for retaining potentially security-relevant flows and ensures that \sysname\ forwards all traffic associated with low-client destinations to the upstream application. Importantly, even under this cautious setting, our experiments show that \sysname\ can still achieve substantial overall traffic reduction while amplifying visibility where it matters most.

\vspace{3pt}
\noindent \textbf{Random Sampling Rates:} 
We set $\tau_{l} = 1/10$ for Locally Popular traffic and $\tau_{g} = 1/100$ for Locally \& Globally Popular traffic. These rates are intentionally conservative: large networks commonly sample such traffic at 1:1000 or lower, but we retain a higher fraction to avoid losing potentially relevant signals while still reducing volume significantly.

\vspace{3pt}
\noindent \textbf{First N Packets Threshold:} 
For traffic categories like {\em Shared Services} and {\em Unidentified}, our approach is to capture the first $N$ packets of each flow. Given that most Internet traffic is now encrypted, critical details related to encrypted traffic, such as the contacted domain (SNI) and server certificates, are often found in those initial packets, whereas packets in the middle of a TLS session offer very little valuable information, especially for upstream applications such as NIDS, which often leverage deep packet inspection.

To decide on the value of the parameter $N$ to be used in our evaluation, we analyzed TLS v1.2 and HTTP traffic from several minutes of real-world traffic, which contained {\em 23,762} TLS connections and {\em 1,074} HTTP connections. For TLS traffic, we counted all packets from the initial TCP SYN until a Server Hello message was observed. For HTTP connections, we counted all packets from the initial TCP SYN packet until the first packet of the first HTTP response was received. Figure~\ref{fig:fist_n_count} shows the distribution of this traffic. Based on these results, we decided to set $N=20$, which allows us to capture the vast majority of initial HTTP request/response headers, TLS SNI and SSL certificates.

\subsection{Live Network Experiments}
\label{sec:real_deployment}
\begin{figure}[!t]
  \centering
  \includegraphics[width=0.8\columnwidth]{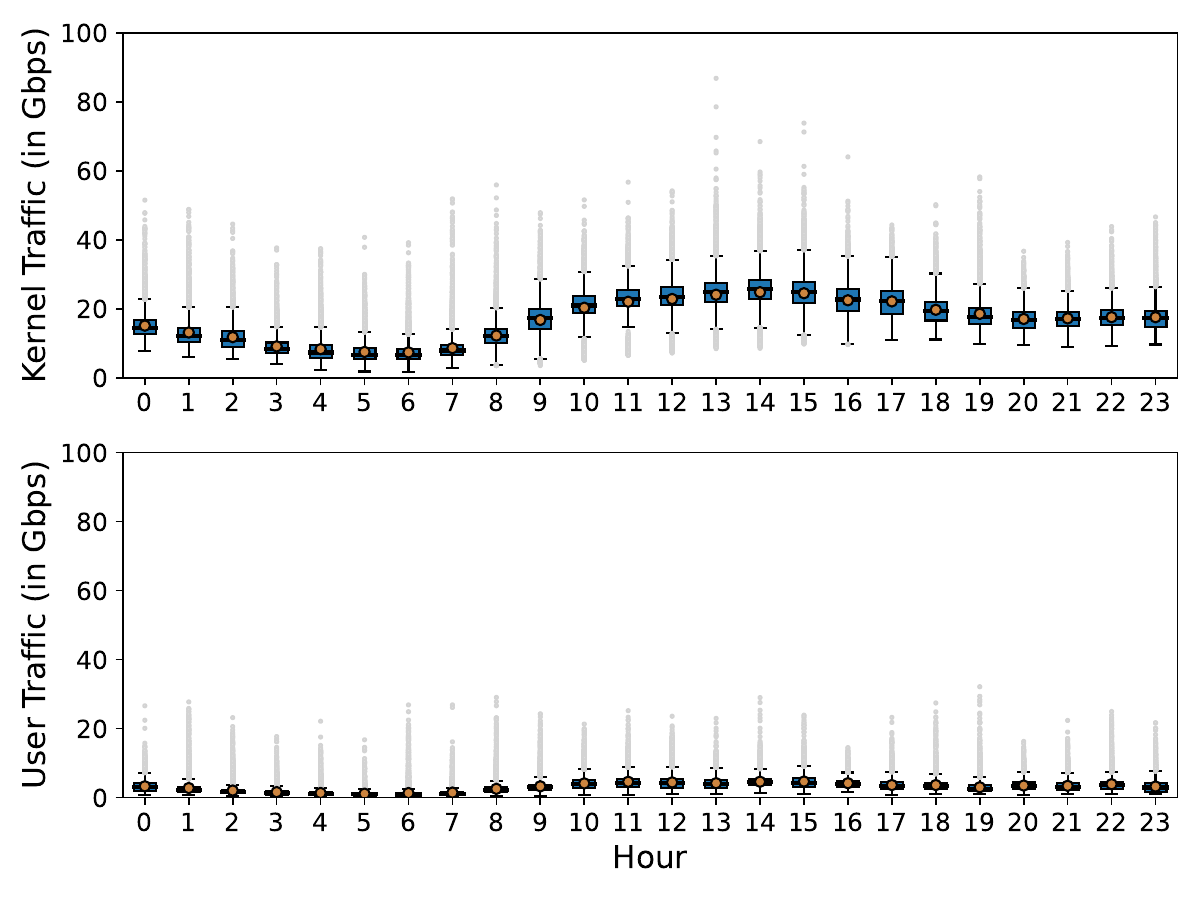}
  \caption{Traffic measurements per hour observed before (top) and after (bottom) \sysname's sampling is applied. }
  \label{fig:kernel_user_traffic}
\end{figure}

We evaluated our system by deploying it on live traffic from a mirrored 100GbE link that connects a large academic network to the Internet. During deployment, we frequently collected traffic measurements (every 10 seconds), including the count of packets, bytes, traffic speed, and average packet size, both overall and per category, that were mirrored to our network interface for a period of {\em 16} days, between {\em 02-14-2025} and {\em 04-04-2025}, with {\em 10} consecutive days and another {\em 6} days sampled within the observation period. The measurements were gathered at the kernel level for each traffic category before and after applying \sysname's sampling policies. Figure~\ref{fig:kernel_user_traffic} (top) shows the distribution of traffic rate for \textit{unsampled} live traffic observed over the 16-day monitoring period, broken down by hour. The bandwidth varies significantly across different times of the day, ranging from an average of 7 Gbps to 25 Gbps and reaching a peak of 87 Gbps during the busy hours of the day, which is typical on a 100 Gbps link, since real networks rarely drive the full theoretical line rate. This fluctuation is expected with live traffic, as it reflects the Internet usage patterns of the connected clients.

When designing our system, one of the primary goals was to considerably reduce traffic while ensuring that the filtered traffic offers an enhanced focus on security-related traffic. Figure~\ref{fig:kernel_user_traffic} (bottom) illustrates the impact of \sysname's sampling on the overall traffic throughout the hours of the day. 
To better represent the overall reduction achieved through \sysname and the contribution of various traffic categories to this reduction,
we performed further evaluations, as detailed below.

\begin{figure}[!t]
  \centering
  \includegraphics[width=0.85\columnwidth]{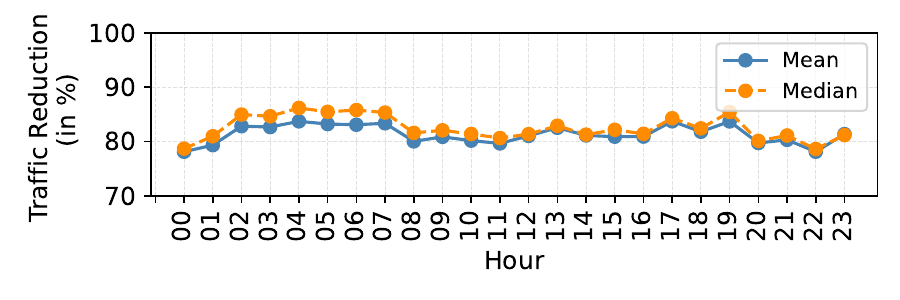}
  \caption{Traffic reduction (in \%) calculated per hour.}
  \label{fig:reduction_percent}
\end{figure}

\begin{figure*}[!t]
  \centering
  \includegraphics[width=0.85\textwidth]{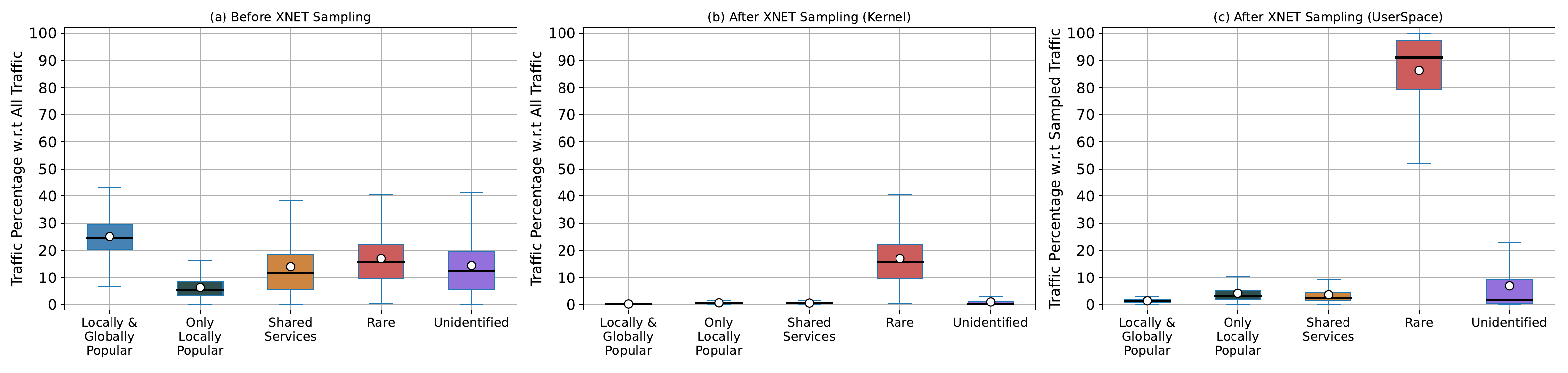} 
  \caption{Distribution of traffic per category. Percentage of each traffic category  a) before \sysname's sampling w.r.t all traffic b) after \sysname's sampling w.r.t all traffic c) after \sysname's sampling w.r.t sampled traffic forwarded to userspace.}
  \label{fig:traffic_tier_percent}
\end{figure*}

\vspace{3pt}
\noindent \textbf{Traffic Reduction Percent:} To evaluate this, we first calculate  {\em Traffic\_Reduction\_Percent}, a measure of the traffic reduction in terms of traffic speed calculated for each 10-second data pair(before and after \sysname's sampling). For each hour, we further aggregate {\em Traffic\_Reduction\_Percent} calculated per 10-second intervals and compute the average and median values as illustrated in Figure~\ref{fig:reduction_percent}. As shown, \sysname achieved an average traffic reduction rate ranging from {\em 78\%} to {\em 84\%}.

\vspace{3pt}
\noindent \textbf{Traffic Reduction Contributors:} Configuration~\ref{lst:config1} (in Appendix) displays thresholds used for traffic categorization and sampling rates applied for each category. These thresholds were chosen based on observations from real traffic as previously explained in section~\ref{sec:threshold_selection}. On applying these policies, we examine the distribution of traffic across traffic categories and their contribution to this reduction. 
Figure~\ref{fig:traffic_tier_percent}(a) depicts the distribution of traffic across the traffic categories before applying \sysname's sampling policies. In this case, the {\em Locally and Globally Popular} category accounts for the highest median traffic percentage, reaching up to {\tt 43\%}. The {\em Unidentified} and {\em SharedServices} categories follow, reaching up to {\tt 41\%} and {\tt 38\%} each. Most importantly, the ``Rare'' category forms less than {\tt 40\%} of the traffic over the observed period with an average of {\tt 15\%}. These numbers further confirm that most traffic in a network is related to popular destinations.

Next, Figure~\ref{fig:traffic_tier_percent}(b) displays the impact in the traffic percentage for each category after \sysname's sampling was applied. From this plot, it is evident that the major category that contributes to the reduction is {\em Locally and Globally Popular} since we sample this category heavily (using 1\% sampling rate) due to its ``low risk'' assessment, leading to a  {\tt 25\%} reduction on average. For categories deemed ``medium risk'', namely {\em SharedServices} and {\em Unidentified}, we collected the first 20 packets per bi-directional flow. As a result, their average reduction contribution to overall traffic is slightly lesser, nearly {\tt $15\%$} on average. On the other hand, since we collect all traffic categorized as {\em Rare}, there is no reduction observed for this category. Note that {\em Only Locally Popular} traffic (sampled at 10\%) is almost negligible in both cases.

Finally, Figure~\ref{fig:traffic_tier_percent}(c) depicts each category's representation in the resulting sampled traffic, which is forwarded to the user space and upstream applications. As mentioned throughout the paper, the main goal of this research is to ``magnify'' the focus on {\em Rare} traffic, which holds significant security value. This plot confirms this phenomenon, with the {\em Rare} traffic category exhibiting a notable increase, with a higher median traffic percentage than all other categories, forming over {\tt 85\%} of the filtered traffic on average. This represents a 5-fold amplification compared to the case where traditional random sampling would have been applied {\tt ($<15\%$)}. Because of the amplification of this traffic category, we expect \sysname's sampling approach to be favorable for upstream security applications such as IDS, for which we provide empirical evidence in section~\ref{sec:suricata}.

\input{performance_primitives}
\input{security_value}

%% file: performance_primitives.tex
\subsection{Ablation Study of XDP Packet Handler}
\label{sec:performance_scaling}
In this evaluation, we dissect the XDP Packet Handler into building blocks of basic operations and evaluate their performance in isolation when exposed to 100Gb/s traffic. This ablation study can reveal potential scalability obstacles and inform decisions about the impact and feasibility of incorporating additional filtering steps in the early packet processing pipeline.

\vspace{3pt}

\noindent \textbf{XDP basic operations:}
As discussed in subsection \ref{subsec:xdp}, the module running at the driver level applies different sampling policies on the incoming packets according to their assigned traffic category. Those tasks can be broken down into a set of basic operations that rely on eBPF maps and functions that can be executed in the restricted kernel environment.  
First, we create simple, self-contained XDP programs to evaluate the following operations in isolation in a controlled environment:

\setlength{\fboxsep}{2pt}

\noindent$\blacktriangleright$\textbf{baseline:} Corresponds to the number of packets reaching the XDP hook when no further processing takes place.

\noindent$\blacktriangleright$\textbf{increaseCounter:} A counter is implemented as an entry of a \url{BPF_MAP_TYPE_PERCPU_ARRAY} and is increased by one for every arriving packet before the packet is dropped. %
Each CPU maintains its own instance to avoid race conditions and synchronization overhead (PERCPU).  

\noindent$\blacktriangleright$\textbf{samplePacket:} A random number is generated for each packet and compared to a predefined sampling rate. It's used to implement random packet sampling for some categories.

\noindent$\blacktriangleright$\textbf{hashLookup:} For each packet, the source IP is extracted and looked up in a map of type \url{BPF_MAP_TYPE_HASH}. It introduces the additional overhead of the lookup operation compared to the baseline case. 

\noindent$\blacktriangleright$\textbf{flowTracking:} A flow key is first created by extracting the (srcIP, dstIP, srcPort, dstPort, proto) tuple from the TCP packets and then looked up in a \url{BPF_MAP_TYPE_PERCPU_LRU_HASH} map. The value is increased by one or initialized to one if missing. It is necessary for keeping only the first $N$ packets from each flow, which typically contain the most relevant security features. Again, notice the PERCPU designation since each CPU has visibility to a unique subset of packets. The LRU implements the eviction policy in case the map reaches capacity.

\noindent$\blacktriangleright$\textbf{AF\_XDPSocket:} Each packet is forwarded to a user-space application through \url{AF_XDP} sockets. Packets arriving at different receive queues (RXQ) are sent to different sockets, to increase parallelism. %

\noindent$\blacktriangleright$\textbf{userWriteDisk:}
Includes sending the packets through \url{AF_XDP} sockets as in the previous case, with the added overhead of each receiving thread buffering the packets and eventually writing them to the disk in PCAP format.

\noindent$\blacktriangleright$\textbf{fullDeployment:} The complete XDP Packet Handler module with packet processing pipeline(refer  Section~\ref{subsec:xdp}) combines all the operations mentioned above and measures the worst-case performance, meaning each packet has to go through all the steps on the kernel side without the option of early exit. 

\begin{figure*}[!ht] %
  \centering
  \captionsetup{skip=2pt}
  \includegraphics[width=0.95\textwidth]{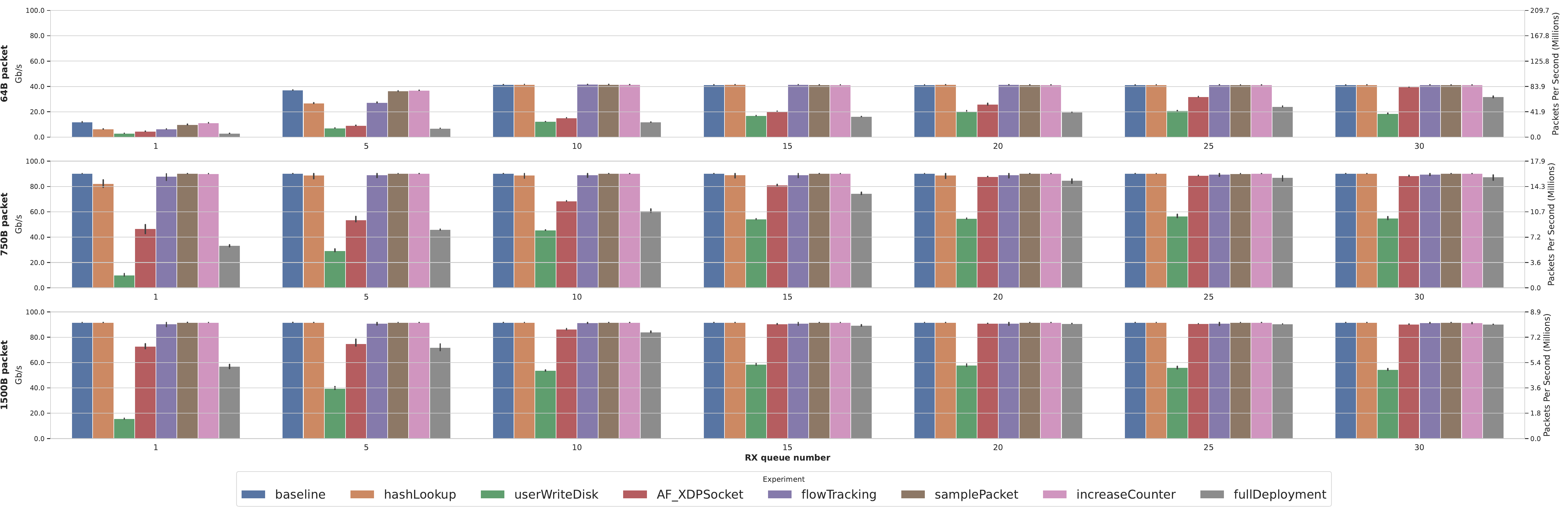} 
  \caption{Performance of basic operations used by the XDP Packet Handler for three packet sizes (64B,750B,1500B). The left vertical axis corresponds to Gb/s, while the right vertical axis is the same value converted to million packets per sec.}
  \label{fig:xdp_primitives}
\end{figure*}

\vspace{3pt}
\noindent \textbf{Testing methodology:} After compiling a simplified XDP program for each basic operation, we stress-test their throughput for different configurations under a saturated 100GbE link. For this task, we utilize Pktgen\cite{pktgen}, a popular open-source, high-performance traffic generator based on DPDK, to replay a large PCAP file at high rates. Pktgen runs on the sibling machine and transmits the packets to the box running \sysname. %
It is well known that the size of the packets impacts the overall packet processing speed for network monitoring systems~\cite{schultz2011passive, ntop_n2disk}. In general, smaller packets have a higher overhead per byte, resulting in a smaller throughput (Gbps), even though the number of packets processed is larger. Therefore, we showcase the behavior of the programs under three different packet sizes: 64B as the smallest size typically used for benchmarking purposes~\cite{schultz2011passive, ntop_n2disk}, 1500B as the typical Ethernet MTU value, and 750B as an intermediate value. For reference, the average packet in our academic network traffic feed is 900B. Aside from the packet size, the second tunable parameter in the experiments is the number of RX (receive) queues. To fully utilize the available RX queues, we carefully select the IP addresses of the packets to send to ensure uniform distribution after Receive Side Scaling (RSS) is applied by the NIC. Each experiment \textit{(packet size, RX queue, operation type)} is repeated 10 times, with each run lasting 120 seconds.

\vspace{3pt}
\noindent \textbf{Performance Results:} Figure \ref{fig:xdp_primitives} shows that the maximum throughput across the 64-byte packet size experiments is significantly lower compared to 750 or 1500. In our setup, Pktgen could not replay those small packets faster than 40Gbps (equivalent to 83 million packets per second). For the 750B and 1500B packet sizes, on the other hand, Pktgen replay speed was around 93\% of the maximum bandwidth. Smaller packet sizes result in higher PPS (packet per second, left y-axis) values but lower Gbps (right y-axis) throughput because of the extra overhead per byte. While 64-byte packets are not a meaningful portion of network traffic, these cases serve to better reveal the nuanced overhead across the various functions that would not be as evident when looking at larger packet sizes (for example, the performance gap between \textit{flowTracking} and \textit{samplePacket} is more apparent when using 1-5 RX queues with 64-byte packets).

In the cases of 750 and 1500-byte packets, 10 to 15 RX queues handle a 100GbE link comfortably for most experiments. As expected, the \textit{fullDeployment} case is the most demanding since it supersets all the previous operations (except the disk writing part, as mentioned). However, even in this case, 20 RXQs appear capable of processing all packets. Sending packets through an \url{AF_XDP} socket causes the most overhead on the kernel's side. Comparatively, additional packet processing, such as retrieving information from eBPF maps, parsing packet headers, or counting packets per flow, makes a minor contribution. Flow tracking and retrieving information from a hash map have comparable costs, since the former is basically updating a hash map with the extra inexpensive steps of extracting the flow key from the packet. This means the system can be extended by incorporating additional or more complex computations without a noticeable performance impact or even a net gain, especially if the extra analysis allows for more packets to be dropped early.

As the only disk-limited operation, \textit{userWriteDisk} has the lowest throughput. In our case, we found that our NVMe setup could handle file writing with higher throughput than reported in the experiments. The reason for this discrepancy comes from the ``wasted" time the threads spend copying over the packets from the buffer to the PCAP file. During this time, new packets cannot be added to the buffer. More complicated solutions that use a rotating set of buffers with additional writing threads could alleviate the problem, but this would increase the complexity of our system with minimum impact in the real-world deployment where only a small fraction of 15-25\% of the packets might need to be stored at any given time, as a result of traffic reduction.

%% file: security_value.tex
\vspace{-5pt}

\subsection{Security Value of the Reduced Traffic}
\label{sec:suricata}

\definecolor{darkgray}{rgb}{0.3, 0.3, 0.3}

\begin{table*}[h!]
\centering
\scriptsize
\begin{tabular}{ |c||c|c|c||c|c|c||c|c|c||c|c|c||c|  }
\hline
\multirow{3}{*}{\makecell{\textbf{Malware} \\ \textbf{Sample ID}}} & \multicolumn{13}{|c|}{\textbf{Traffic Categories}} \\
\cline{2-14}
& \multicolumn{3}{c||}{\textbf{DNS}} & \multicolumn{3}{c||}{\textbf{Popular}} & \multicolumn{3}{c||}{\textbf{Shared Services}} & \multicolumn{3}{c||}{\textbf{Rare}} & \textbf{Other} \\
\cline{2-14}
& \textbf{M} & \textbf{S(M)} & \textbf{X(S(M))} & \textbf{M} & \textbf{S(M)} & \textbf{X(S(M))}& \textbf{M} & \textbf{S(M)} & \textbf{X(S(M))}  & \textbf{M} & \textbf{S(M)} & \textbf{X(S(M))}  & \textbf{M} \\
\hline
\hline
sample6 & 3724 & 6 & 6 & 298 & 0 & 0 & 2856 & 4 & \cellcolor{gray!50}3 & 48674 & 6 & 6 & 310  \\
\hline
sample10 & 145 & 1 & 1 & 315 & 9 & \cellcolor{gray!50}0 & 1551 & 1 & 1 & 6230 & 261 & 261 & 1661\\
\hline
sample12 & 829 & 1 & 1 & 54 & 0 & 0  & 6381 & 4 & \cellcolor{gray!50}3 & 121983 & 10 & 10 & 239 \\
\hline
sample20 & 163 & 5 & 5 & 233 & 1 & \cellcolor{gray!50}0  & 8377 & 69 & 69 & 7488 & 61 & 61 & 68 \\
\hline
\hline
\textbf{All 20 Samples} & 14605 & 54 & 54 & 2235 & 10 & 0 & 46569 & 255 & 253 & 345448 & 2473 & 2473 & 17557 \\
\hline
\end{tabular}
\caption{Distribution of packets from various malware samples into the traffic categories we use. \textbf{M}=packets per malware sample, \textbf{S(M)}= the Suricata alerts that $M$ generated and \textbf{X(S(M))} the alerts after \sysname is deployed. Marked gray are the cases our system misses an alert-related packet. The column ``Other" consists of internal or non-IPv4 traffic.}
\label{tab:malware_samples_info}
\end{table*}

This section demonstrates the security value of the selectively reduced traffic generated by \sysname. For our evaluation, we use Suricata, a popular upstream security application with intrusion detection (IDS), intrusion prevention (IPS), and network monitoring capabilities. Suricata analyzes raw packets and generates alerts when suspicious activity or known threats are identified. For this experiment, we focus on alerts that originate from the {\em Emerging Threat} (ET) ruleset. The ET ruleset is a community-driven and regularly updated detection signatures (rules) database that helps identify various network threats, such as C2 communication, exploits, network anomalies, scanning, etc. 

\vspace{1pt}
\noindent \textbf{Dataset:} The first step is to create a dataset with security-relevant traffic instances. For that purpose, we downloaded 20 PCAP files with verified malicious traces from \url{www.malware-traffic-analysis.net} and merged them (full list in Appendix \ref{app:malware_pcap_list}). In total, the malware traces included {\em 426,414} packets that we denoted as $M$. Next, we uniformly mix $M$ with traffic captured from our institution's feed, $B$, whose purpose is to conceal the presence of malicious traces ($M$). $B$ consists of traffic from 800 local clients over a 15-minute window. The constraint on the number of clients and the capture duration is due to the limited PCAP size we can replay with Pktgen, as it incurs heavy memory overhead per packet. The resulting mixed traffic, $T = M \cup B$, includes {\em 2,664,255} packets. We use Suricata to extract the alerts from our traffic mix, annotating $S(M)$ and $S(B)$ the portion of the alerts due to packets in $M$ and $B$ respectively. Not surprisingly, $S(B)=0$ as our institution traffic can be considered mostly benign, with a trivial percentage of malicious traffic.
On the other hand, $S(M)=2,792$ is the number of alerts from the malicious traces. The exhaustive list of the unique signature descriptions for these alerts can be found in Appendix \ref{app:alerts}. 

Next, we replayed $T$ at the highest possible rate with Pktgen and used \sysname to extract the reduced filtered traffic $X(T)$. \sysname was deployed with the internal state (IP-Category mappings) it gathered until the moment our institution's traffic ($B$) was captured. \sysname achieved an overall reduction rate of $75\%$ on the mixed traffic with perfect packet visibility, even at the highest replay speed. Specifically, $X(M)=373,586$, or $88\%$  packets from the original malware traces $M$ were retained. In terms of $X(B)$, only $13\%$ of normal traffic survived post-sampling, in accordance with the overall behavior we observed in Section \ref{sec:real_deployment}. Our system significantly reduced "normal" traffic originating from networks it was already familiar with and for which it possessed accurate IP-to-category intelligence. In contrast, the reduction was considerably lower for the new traffic patterns introduced with $M$. 

To assess the value of our reduced traffic, we fed it to Suricata. Suricata detected $S(X(M))=2,780$ alerts from the portion of the malicious traffic that survived our filtering. That is {\tt 99.6\%} of the original $S(M)=2792$ alerts. Detailed results on the detection for a subset of malware samples are presented in Table~\ref{tab:malware_samples_info}, while the results for all malware samples can be found in Table~\ref{tab:malware_samples_info_full} that is moved to the Appendix due to space constraints. The table also shows the distribution of the packets from the malicious samples over our defined categories.

As expected, we found that packets in the Rare Category triggered most alerts. These destinations were previously unseen or accessed by only a few clients in our university network. These alerts, along with all DNS-related alerts, passed through our filtering process because our policy retains every DNS and Rare packet. Four cases where our system dropped certain packets of interest, leading to fewer alerts than the baseline, are highlighted in gray.
\vspace{1pt}

\noindent \textbf{Missed Alerts analysis:} Overall, {\em 255} packets belonging to the {\em Shared Services Traffic Category} were responsible for alerts, and our system only missed two cases. On closer inspection, the reason was that those alerts were triggered on payload analysis of plain-text HTTP traffic, and since we only keep the first {\em 15} packets per connection, some payload information was omitted. Notably, although it missed some packets containing the payload, the system successfully sampled the first 15 packets from the same flow, which also generated relevant alerts, ensuring the client would be flagged despite the missed packets.  Finally, in two cases, {\em 10} alerts were due to traffic exchanged with {\em Globally Popular} destinations that we heavily sample, hence we missed those. In the case of {\tt sample10}, all {\em 9} alerts were related to one external IP that we labeled as popular 
and had the signature: {\tt ET INFO Observed Telegram Domain (t .me in TLS SNI)}. The other case of a popular destination generating an alert is {\tt sample20} with signature: {\tt ET MALWARE FormBook CnC Checkin (GET) M5}. The IP under investigation had been associated with \url{github.com} and \url{github.io}. After looking up both IPs in VirusTotal, we got {\em 10} and {\em 16} entities that flagged them as malicious, respectively. Note that these can be easily mitigated by adding a category of malicious IPs as discussed in Section~\ref{subsec:extensions}.

We repeated the same experiment using Snort as the security application, instead of Suricata, and observed almost identical behavior. The results showed a {\tt 99.5\%} alert retention rate after XNET's traffic reduction. Finally, we verified that XNET does not introduce any false positives, meaning that when XNET filters real-world traffic, no new security-relevant alerts are generated with either Suricata or Snort, compared to the unfiltered baseline.

\vspace{3pt}
\noindent \textbf{\sysname vs Alternative Traffic Sampling Approaches.}
We repeated the same steps as in Section \ref{sec:suricata}, replacing \sysname with basic sampling approaches commonly mentioned in the literature \cite{jazi2017detecting}. The sampling techniques we considered and the aggregated results of the experiments (repeated 1000 times) are listed in Table~\ref{tab:xnet_vs_sampling}. The sampling probabilities for each scenario were tuned to achieve the same overall traffic reduction effect as \sysname, namely 75\%.

\begin{table}[t!]
\centering
\scriptsize
\begin{tabular}{ |l|l|c| }
\hline 
\multicolumn{1}{|c}{\textbf{\begin{tabular}[|c|]{@{}c@{}}Sampling\\ Approach \end{tabular}}}                    
& \multicolumn{1}{|c|}{\textbf{\begin{tabular}[|c|]{@{}c@{}}Sampling\\ Description \end{tabular}}}                       & \multicolumn{1}{c|}{\textbf{\begin{tabular}[|c|]{@{}c@{}}Avg. Alert \\ Percentage\end{tabular}}} \\
\hline
\hline
\begin{tabular}[l]{@{}l@{}}Random Packet  Sampling\end{tabular} & Sample 1 out of N packets                                       & 1.2\%                                                           \\ \hline
\begin{tabular}[l]{@{}l@{}}Random Flow \\ Sampling\end{tabular}   & \begin{tabular}[l]{@{}l@{}}Sample all packets from \\ 1 out of N flows\end{tabular}                      
& 21\%                                                            \\ \hline
Sample and Hold                                                  %
 & \begin{tabular}[l]{@{}l@{}}For each randomly sampled packet \\ capture the rest in the same flow 
\end{tabular} 
& 2.9\%    
\\                                    
\hline
\textbf{XNET} & Custom sampling (Section~\ref{sec:traffic_categories}) & \textbf{99.6\%} \\
\hline
\end{tabular}
\caption{XNET sampling vs other sampling approaches.}
\label{tab:xnet_vs_sampling}

\end{table}

Compared to \sysname's performance, all these approaches significantly underperformed, capturing 1.2\%, 2.9\%, and 21\% alerts, depending on the methodology. Random Packet Sampling and Sample-and-Hold struggled the most because many Suricata alerts rely on prerequisites like an established flow or earlier packet observations to trigger, which are unlikely to be captured with Random Sampling. In Sample-and-Hold, the initial packets of each flow are likely to be missed; hence, its performance is only marginally improved.

The Random Flow Sampling results further highlight the importance of capturing the early packet sequence. Although this method still fell short of \sysname’s performance, retaining all packets for a randomly selected set of flows improved the detection rates significantly. This supports the importance of our sampling policy of capturing the first $N$ packets. As none of the aforementioned approaches categorizes traffic, they allocated large portions of their sampling budget to capturing ``normal" traffic and missed most security-relevant events.

\vspace{1pt}
\noindent \textbf{Summary:} From this experiment, we showed that \sysname can achieve significant traffic reduction (75\%) while still retaining more than 99\%  of the security-relevant packets without introducing false positive alerts. This directly translates to gains for upstream applications, for example, packet brokers \cite{arkime} with limited available storage and IDS systems that experience degraded performance as packet rates increase \cite{HU2020IDS100Gbps,waleed2022open}. 

%% file: discussion.tex
\section{Discussion}

\noindent \textbf{Network vs Host Level Monitoring: } We acknowledge that traffic to popular destinations may also conceal malicious activities, as noted in prior works. For instance, malware may use benign services (e.g., Dropbox, Twitter, Slack) for malicious purposes, such as C2 communication and data exfiltration~\cite{wang2022deepc2,iranianslack,russiandropbox,yao2023hiding}. While we do not neglect the severity of such incidents, according to previous studies~\cite{milajerdi2019holmes, alsaheel2021atlas, ulusoy2015guardmr,ji2017rain} such threats are better dealt with at the host or server level, through proper system-level monitoring tools. Also, most traffic to popular Internet services is encrypted and cannot be inspected by traditional NIDS. While SSL man-in-the-middle proxies can be used to block certain types of malicious traffic, these proxies complement but do not substitute other network security appliances, like NIDS and traffic storage systems, which represent our target upstream applications. Furthermore, by abusing legitimate services, adversaries risk rapid forensic attribution, takedown, and criminal prosecution, as those services continuously investigate logs and metadata around their users' activity on the server side. Therefore, our strategy focuses on ``amplifying'' visibility on traffic categories that tend to have higher value for our target upstream security applications while more heavily sampling traffic towards popular destinations.

Notably, XNET does not heavily downsample all traffic associated with popular destinations; instead, it explicitly isolates CDN and cloud-hosted infrastructure into the Shared Services category and applies a more conservative first-N-packets policy, preserving the initial packets where most attack-relevant information resides.

\vspace{3pt}
\label{subsec:extensions}
\noindent \textbf{Extending \sysname.} 
As mentioned earlier in the paper (Section~\ref{sec:traffic_tiers}), we defined an example set of traffic categories to illustrate the effectiveness of our system. At the same time, \sysname's traffic categories can be easily extended to adapt the system to different network environments and network administrators' preferences. For instance, consider the Shared Services category. Some studies have suggested that some CDNs or cloud services can be more prone to abuse than others~\cite{Subramani2024CDNFronting,Tandon2020CloudMisBehavior,Alrwais2017BPH}. Based on these observations, a network administrator may want to further divide the Shared Services into sub-categories, and assign more fine-grained traffic sampling policies to sub-categories that is historically associated with higher levels of abuse. 

Another potential category is ``Malicious'' traffic. This new category could be easily included by simply adding malicious or low-reputation IP addresses with a new label tag to the IP-Category eBPF map used by the XDP Packet Handler, and by associating a sampling policy (e.g., No Sampling) to this new category. For example, one may use IP reputation systems like VisrusTotal~\cite{virustotal} or Cisco's Talos~\cite{talos}. However, because \sysname is not designed to be a detection tool and we have limited access to the VirusTotal platform (insufficient for labeling the large number of external IPs found in live traffic), we did not implement this category. At the same time, our analysis of missed alerts in Section~\ref{sec:suricata} suggests that including such a new Malicious category would have allowed us to avoid all those missed alerts. On further investigation, we found that {\tt 0.6\%} or {\em 57} IPs with a Globally Popular label are flagged as malicious in VirusTotal by more than three vendors (the typical threshold researchers use). Notably, traffic to these flagged IPs constituted only {\tt $0.5\%$} of all Globally Popular traffic, which would have caused a rather small increase in traffic retained by \sysname.

%% file: related_work.tex
\section{Related Work}

\vspace{3pt}
\noindent \textbf{Traffic Sampling and Filtering.}
A few works exist that propose customizable sampling approaches. Studies~\cite{ramachandran2008fast,Zhang2011AdaptiveSampling} present sampling geared towards collecting specific subpopulations, boosting detection of botnet-related attacks. In ~\cite{jazi2017detecting}, the authors study the effect of different sampling approaches on detecting application layer DoS attacks.  \cite{papadogiannakis2013scap} focuses on reassembling traffic flows and amplifying visibility for short and recent packet streams. \cite{wan2022retina} is designed to apply multilayer packet filters to answer a specific query, discarding irrelevant packets. Traffic analysis and capture platforms such as \cite{takano2015sf,morariu2008dicap,gad2015monitoring} rely on deployment across multiple servers to scale. Unlike these approaches, \sysname focuses on increasing traffic visibility towards ``Rare" destinations that may hold traffic related to a wide range of malicious activities and adapts according to external and internal (network's behavior) intelligence.

\vspace{3pt}
\noindent\textbf{Hardware-Accelerated Approaches.} Recent advances in programmable data-plane technology have driven increased interest in hardware-based solutions. While programmable switches can process traffic at rates up to Tb/s, they are constrained by limited hardware resources, which restricts their ability to perform complex computations compared to a general-purpose server. As a result, existing solutions primarily target specific tasks such as DDoS detection \cite{zhang2020poseidon, liu2021jaqen, liu2016one}, heavy-hitter detection \cite{sivaraman2017heavy}, or other specialized functions \cite{barradas2021flowlens,sonchack2018turboflow}. For more complex analysis, these systems often rely on close integration with external servers \cite{panda2021smartwatch}. SmartNICs have also been employed to support hardware-accelerated IDS applications \cite{zhao2020achieving}. However, hardware-specific languages like P4 introduce additional challenges \cite{sonchack2021lucid} that make application development and maintenance more complex.

%% file: conclusion.tex
\section{Conclusion}

In this paper, we presented \sysname, a dynamic security-aware traffic categorization and sampling system designed to run on commodity hardware. We explained how the two primary components, the XDP Packet handler and the Traffic Profiler, collaborate to assign categories to individual packets, sample them accordingly, and adapt the classification based on external data feeds and local network intelligence.  We evaluated \sysname under a real operational scenario of a large academic institution demonstrating a packet reduction rate of 78-84\%. We also performed controlled experiments to show that this reduction can amplify the visibility of security-relevant events, which can be extremely beneficial for upstream applications that struggle with high packet rates. Finally, stress tests of the packet processing components of our system showcase its ability to scale and operate under 100Gb/s traffic conditions.

%% file: appendix.tex
\appendices

\input{ethical}

\section{XDP Compared to DPDK} 
\label{app:design_choices}

To facilitate packet processing at line rate, \sysname leverages the capabilities of XDP instead of opting for other technologies. Although frameworks such as DPDK can offer similar or even greater performance enhancements, XDP offers distinct advantages due to its direct integration with  Linux kernels, enabling low-latency packet processing within the kernel's networking stack. Unlike DPDK, a user-space solution that completely bypasses the kernel to maximize throughput, XDP uses the kernel’s efficiency while minimizing overhead. Performing packet filtering and processing at the driver level reduces the cost of handling packets and avoids excessive interrupt handling and context switching. Moreover, DPDK requires significant development effort and encounters compatibility issues across kernel versions and hardware platforms.  Its lack of convenient kernel-level packet-processing libraries further complicates the coding process, as developers must manage low-level details manually, making development more time-consuming and prone to errors. In contrast, XDP, introduced with Linux kernel 4.8, is widely portable across hardware architectures and supported by most network drivers. It provides developers with a familiar and flexible environment without domain-specific languages, restrictive pipelines, or error-prone and lengthy development processes. Finally, some network cards even allow offloading XDP programs to the hardware instead of the device's driver \cite{netronomeoffload}, potentially bridging any performance gap.

To summarize, XDP offers the following benefits: 1) early-stage packet filtering (at the network driver level) before packets reach the kernel stack, reducing latency and improving performance 2) not restricted by specialized hardware thus avoiding resource limitations typical of hardware solutions 3) its integration with eBPF's safety features, such as verification and sandboxing, ensure that the program does not compromise the stability or security of the overall system. 

\section{\sysname's Limitations}
Our framework leverages XDP for fast packet filtering at line rates up to 100Gb/s. This limits its deployment to Linux distributions. Additionally, support for XDP and AF\_XDP more specifically was introduced in kernel versions 4.8 (released in 2016) and 4.18 (released 2018) respectively and has expanded upon since. To harness most of the performance gains, XDP must be natively supported by the NIC driver. In practice we found this to be the case for the majority of the popular NICs. 

We process only IPv4 traffic ignoring other protocols. The same framework can work with IPv6 traffic with minor changes, such as making the map keys larger to accommodate 64 bit addresses and adding some additional code to handle the packet accordingly. We focused on IPv4 to make development simpler. In our academic network, the portion of IPv6 traffic is approximately 20\%.

\section{Unique Suricata Alerts}
\label{app:alerts}
\scriptsize
{
\begin{itemize}
\item ET DNS Query for .cc TLD
\item ET DNS Query to a *.top domain - Likely Hostile
\item ET DROP Spamhaus DROP Listed Traffic Inbound group 23
\item ET DROP Spamhaus DROP Listed Traffic Inbound group 25
\item ET DROP Spamhaus DROP Listed Traffic Inbound group 30
\item ET DROP Spamhaus DROP Listed Traffic Inbound group 33
\item ET EXPLOIT SUSPICIOUS Possible CVE-2017-0199 IE7/NoCookie/Referer HTA dl
\item ET HUNTING curl in DNS TXT Response
\item ET HUNTING curl User-Agent to Dotted Quad
\item ET HUNTING EXE extension in DNS TXT Response
\item ET HUNTING GENERIC SUSPICIOUS POST to Dotted Quad with Fake Browser 1
\item ET HUNTING PDF extension in DNS TXT Response
\item ET HUNTING Successful PROPFIND Response for Application Media Type
\item ET HUNTING SUSPICIOUS Dotted Quad Host MZ Response
\item ET HUNTING WebDAV Retrieving .zip
\item ET INFO Dotted Quad Host DLL Request
\item ET INFO Dotted Quad Host ZIP Request
\item ET INFO Executable Download from dotted-quad Host
\item ET INFO Executable Retrieved With Minimal HTTP Headers - Potential Second Stage Download
\item ET INFO EXE - Served Attached HTTP
\item ET INFO External IP Address Lookup Domain (ipify .org) in TLS SNI
\item ET INFO External IP Lookup Domain (ipify .org) in DNS Lookup
\item ET INFO Free Hosting Domain (*.freehostia .com in DNS Lookup)
\item ET INFO HTTP Request to a *.top domain
\item ET INFO HTTP traffic on port 443 (POST)
\item ET INFO Observed DNS Query to .biz TLD
\item ET INFO Observed Telegram Domain (t .me in TLS SNI)
\item ET INFO Observed ZeroSSL SSL/TLS Certificate
\item ET INFO OpenSSL Demo CA - Internet Widgits Pty (O)
\item ET INFO Packed Executable Download
\item ET INFO PE EXE or DLL Windows file download HTTP
\item ET INFO Possible HTA Application Download
\item ET INFO PS1 Powershell File Request
\item ET INFO TLS Handshake Failure
\item ET INFO Windows Powershell User-Agent Usage
\item ET JA3 Hash - [Abuse.ch] Possible Dridex
\item ET MALWARE [ANY.RUN] DarkGate Check-In HTTP Header (POST)
\item ET MALWARE BackConnect CnC Activity (Set Sleep Timer)
\item ET MALWARE BackConnect CnC Activity (Start VNC) M1
\item ET MALWARE Darkgate Stealer CnC Checkin (POST) M2
\item ET MALWARE DNS Query to Darkgate Domain (saintelzearlava .com)
\item ET MALWARE DNS Query to Darkgate Domain (trans1ategooglecom .com)
\item ET MALWARE DNS Query to IcedID Domain (brojizuza .com)
\item ET MALWARE DNS Query to IcedID Domain (manjuskploman .com)
\item ET MALWARE DNS Query to IcedID Domain (qousahaff .com)
\item ET MALWARE FormBook CnC Checkin (GET) M5
\item ET MALWARE IcedID CnC Domain in DNS Lookup (aptekoagraliy .com)
\item ET MALWARE IcedID CnC Domain in DNS Lookup (joekairbos .com)
\item ET MALWARE IcedID CnC Domain in DNS Lookup (seedkraproboy .com)
\item ET MALWARE Observed IcedID Domain (asleytomafa .com in TLS SNI)
\item ET MALWARE Observed IcedID Domain (brojizuza .com in TLS SNI)
\item ET MALWARE Observed IcedID Domain in DNS Lookup (spkdeutshnewsupp .com)
\item ET MALWARE Observed IcedID Domain (manjuskploman .com in TLS SNI)
\item ET MALWARE Observed IcedID Domain (qousahaff .com in TLS SNI)
\item ET MALWARE PowerShell Script Downloading Emotet DLL
\item ET MALWARE Terse alphanumeric executable downloader high likelihood of being hostile
\item ET MALWARE W32.DarkVNC Variant Checkin
\item ET MALWARE WebDAV Retrieving .zip from .url M1 (CVE-2023-36025)
\item ET MALWARE WebDAV Retrieving .zip from .url M2 (CVE-2023-36025)
\item ET MALWARE Win32/Emotet HTML Template Response
\item ET MALWARE Win32/IcedID Request Cookie
\item ET MALWARE Win32/IcedID Requesting Encoded Binary M4
\item ET MALWARE Win32/SSLoad Payload Request (GET)
\item ET MALWARE Win32/SSLoad Payload Response
\item ET MALWARE Win32/SSLoad Registration Activity (POST)
\item ET MALWARE Win32/SSLoad Registration Response
\item ET MALWARE Win32/SSLoad Tasking Request (POST)
\item ET MALWARE Win32/SSLoad Tasking Response
\item ET MALWARE Windows Executable Downloaded With Image Content-Type Header
\item ET \url{WEB_CLIENT} WebDAV GET Request for .url Flowbit Set
\item ET \url{WEB_CLIENT} WebDAV Retrieving an .url
\end{itemize}
}

\begin{minipage}[t]{.9\linewidth}
\section{Category and Threshold Configuration}
\label{app:configuration}
An example of the configuration file that defines the categories and the thresholds that we use throughout the experiments in this paper is presented in Configuration \ref{lst:config1}. Notice that the operator can easily change the various thresholds and define new categories as required, further customizing \sysname for their needs.

\begin{lstlisting}[{caption={Sampling rates and thresholds.}, label={lst:config1}}]
  {
    "Locally_and_Globally_Popular": {
      "sampling_rate": "1%",
      "retention_period_in_days": 5,
      "threshold": {
        "IP_rank": "<= 20k",
        "local_clients": ">= 100"
      }
    },
    "Locally_Popular": {
      "sampling_percentage": "10%",
      "retention_period_in_days": 3,
      "threshold": {
        "local_clients": ">= 100"
      }
    },
    "Rare": {
      "sampling_percentage": "100%",
      "retention_period_in_days": 0,
      "threshold": {
        "local_clients": "<= 15",
        "total_packets_per_IP": "< 1M"
      }
    },
    "SharedServices": {
      "sampled_packets": "first 20 packets per flow",
      "retention_period_in_days": 1
    }
    "UnIdentified": {
      "sampled_packets": "first 20 packets per flow",
      "retention_period_in_days": 0
    }
  }
  \end{lstlisting}
\end{minipage}

\section{Addditional Results}
\label{app:malware_pcap_list}

 \begin{figure}[!ht]
      \centering
\includegraphics[width=\linewidth]{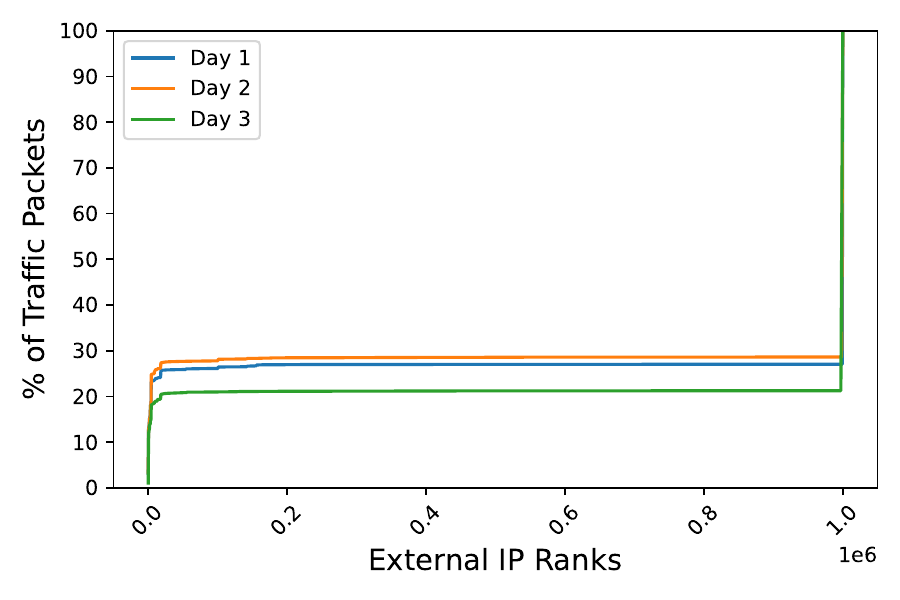}
      \caption{CDF of total traffic per IP rank.}
      \label{fig:cdfrankcount}
  \end{figure}

\definecolor{darkgray}{rgb}{0.3, 0.3, 0.3}

\begin{table*}[ht!]
    \centering
    \begin{tabular}{ |c||c|c|c||c|c|c||c|c|c||c|c|c||c|  }
        \hline
        \multirow{3}{*}{\makecell{\textbf{Malware} \\ \textbf{Sample ID}}} & \multicolumn{13}{|c|}{\textbf{Traffic Categories}} \\
        \cline{2-14}
        & \multicolumn{3}{c||}{\textbf{DNS}} & \multicolumn{3}{c||}{\textbf{Popular}} & \multicolumn{3}{c||}{\textbf{Shared Services}} & \multicolumn{3}{c||}{\textbf{Rare}} & \textbf{Other} \\
        \cline{2-14}
        
        & \textbf{M} & \textbf{S(M)} & \textbf{X(S(M))} & \textbf{M} & \textbf{S(M)} & \textbf{X(S(M))}& \textbf{M} & \textbf{S(M)} & \textbf{X(S(M))}  & \textbf{M} & \textbf{S(M)} & \textbf{X(S(M))}  & \textbf{M}  \\
        \hline
sample1 & 40 & 3 & 3 & 40 & 0 & 0 & 4293 & 163 & 163 & 23 & 0 & 0 & 382\\    
\hline
sample2 & 190 & 14 & 14 & 36 & 0 & 0 & 5671 & 10 & 10 & 2422 & 48 & 48 & 2292  \\
\hline
sample3 & 62 & 4 & 4 & 47 & 0 & 0 & 3552 & 2 & 2 & 2792 & 24 & 24 & 65 \\
\hline
sample4 & 486 & 0 & 0 & 202 & 0 & 0 & 3535 & 0 & 0 & 15740 & 642 & 642 & 5796  \\
\hline
sample5 & 8260 & 9 & 9 & 402 & 0 & 0 & 4433 & 0 & 0 & 74933 & 123 & 123 & 583 \\
\hline
sample6 & 3724 & 6 & 6 & 298 & 0 & 0 & 2856 & 4 & \cellcolor{gray!50}3 & 48674 & 6 & 6 & 310  \\
\hline
sample7 &  96 & 0 & 0 & 0 & 0 & 0 & 362 & 0 & 0 & 8414 & 19 & 19 & 730  \\
\hline
sample8 & 8 & 0 & 0 & 0 & 0 & 0 & 26 & 0 & 0 & 10516 & 177 & 177 & 19 \\
\hline
sample9 & 155 & 0 & 0 & 347 & 0 & 0 & 2084 & 0 & 0 & 8756 & 61 & 61 & 1287  \\
\hline
sample10 & 145 & 1 & 1 & 315 & 9 & \cellcolor{gray!50}0 & 1551 & 1 & 1 & 6230 & 261 & 261 & 1661\\
\hline
sample11 & 10 & 0 & 0 & 0 & 0 & 0  & 96 & 0 & 0 & 8835 & 687 & 678 & 1020 \\
\hline
sample12 & 829 & 1 & 1 & 54 & 0 & 0  & 6381 & 4 & \cellcolor{gray!50}3 & 121983 & 10 & 10 & 239 \\
\hline
sample13 & 146 & 0 & 0 & 94 & 0 & 0  & 832 & 0 & 0 & 4213 & 123 & 123 & 1295  \\
\hline
sample14 & 4 & 0 & 0 & 0 & 0 & 0 & 0 & 0 & 0 & 2562 & 39 & 39 & 0 \\
\hline
sample15 & 22 & 11 & 11 & 0 & 0 & 0 & 0 & 0 & 0 & 3583 & 54 & 54 & 0  \\
\hline
sample16 & 6 & 0 & 0 & 0 & 0 & 0 & 0 & 0  & 0 & 7185 & 47 & 47 & 0  \\
\hline
sample17 & 8 & 0 & 0 & 0 & 0 & 0 & 26 & 0 & 0 & 2836 & 33 & 33 & 0  \\
\hline
sample18 & 194 & 0 & 0 & 10 & 0 & 0  & 2160 & 2 & 2 & 5082 & 34 & 34 & 1726 \\
\hline
sample19 & 57 & 0 & 0 & 61 & 0 & 0  & 334 & 0 & 0 & 3181 & 24 & 24 & 84 \\
\hline
sample20 & 163 & 5 & 5 & 233 & 1 & \cellcolor{gray!50}0  & 8377 & 69 & 69 & 7488 & 61 & 61 & 68 \\
\hline
        
        \hline
        \textbf{Total} & 14605 & 54 & 54 & 2235 & 10 & 0 & 46569 & 255 & 253 & 345448 & 2473 & 2473 & 17557 \\
        \hline

    \end{tabular}
    \caption{Distribution of packets from various malware samples into the traffic categories we use. \textbf{M}=packets per malware sample, \textbf{S(M)}= the Suricata alerts that $M$ generated and \textbf{X(S(M))} the alerts after \sysname is deployed. Marked in gray are the cases where our system misses an alert-related packet. The column ``Other" consists of internal or non-IPv4 traffic.}
    \label{tab:malware_samples_info_full}
\end{table*}

\begin{table*}[th!]
  \centering
  \begin{tabular}{p{0.2\linewidth} | p{0.75\linewidth}}
  \hline
  \textbf{Malware Sample ID} & \textbf{Filename} \\
  \hline
    sample1 & 2023-12-07-DarkGate-infection.pcap \\
\hline
 sample2 & 2023-10-16-IcedID-infection.pcap\\
\hline
 sample3 & 2023-10-31-IcedID-infection-traffic.pcap\\
\hline
 sample4  & 2022-11-17-Bumblebee-infection-traffic.pcap\\
\hline
 sample5 &  2022-01-20-Emotet-epoch5-infection-with-spambot-activity.pcap\\
\hline
 sample6 &  2022-01-20-Emotet-epoch4-infection-with-spambot-activity.pcap\\
\hline
 sample7 &  2022-05-23-IcedID-infection-with-BackConnect-and-Anubis-VNC.pcap\\
\hline
 sample8 &  2023-10-25-DarkGate-infection-traffic.pcap\\
\hline
 sample9  & 2023-10-17-TA577-Pikabot-infection-with-Cobalt-Strike.pcap\\
\hline
 sample10  & 2024-04-18-SSLoad-with-follow-up-Cobalt-Strike-DLL.pcap\\
\hline
 sample11  & 2023-11-20-DarkGate-infection-traffic.pcap\\
\hline
 sample12  & 2022-11-07-part-2-of-2-Emotet-post-infection-with-IcedID-and-Bumblebee.pcap\\
\hline
 sample13  & 2022-12-07-Bumblebee-infection-with-Cobalt-Strike.pcap\\
\hline
 sample14 &  2022-10-10-Qakbot-infection-with-Cobalt-Strike-carved-and-sanitized.pcap\\
\hline
 sample15 &  2023-11-30-DarkGate-infection-traffic.pcap\\
\hline
 sample16 &  2024-01-25-DarkGate-infection-traffic.pcap\\
\hline
 sample17  & 2024-01-30-DarkGate-infection-traffic.pcap\\
\hline
 sample18 &  2022-12-20-IcedID-infection-with-Cobalt-Strike.pcap\\
\hline
 sample19 &  2024-04-04-Koi-Loader-Stealer-infection-traffic.pcap\\
\hline
 sample20  & 2024-08-12-XLoader-Formbook-infection-traffic.pcap \\
\hline
  \end{tabular}
      \caption{Description of Malware Samples in the Dataset}
\end{table*}

%% file: ethical.tex
\section{Ethical Considerations}
To conduct our experiments, we use mirrored traffic from our institution to evaluate our \sysname system. For this, we have sought and obtained approval from our institution's IRB, legal, and cyber teams. Our deployment was categorized as low-risk because even though we were receiving full raw packets, the information of interest was contained in the packet headers. With the exception of DNS packets, where we parse the payload, we did not analyze and did not preserve any information about payload content. While our system can be configured to store raw packets if the network administrator desires, in our experiments, packets were automatically deleted soon after being captured, and this very short-term storage capability was only used to measure the storage requirements for our traffic mix and to collect other high-level statistics. No data from raw traffic were allowed to leave the server, and access to the machine was limited only to authorized researchers. Additionally, we had no access to any information that may be used to attempt mapping IP addresses to network users (e.g., authentication servers or DHCP logs).

\section{Data Availability}
We will release XNET's source code, including all XDP/eBPF components, configuration scripts, and evaluation tools, in an open-source repository upon publication. However, the network traffic datasets used in our experiments cannot be publicly shared due to institutional privacy requirements and the presence of sensitive user information. To protect the privacy of individuals and comply with organizational policies, the raw packet traces and derived statistics from our deployment will remain confidential. We will, however, provide detailed documentation and synthetic examples that enable others to reproduce our methodology without exposing private data.